\documentclass[aps,prb,superscriptaddress,preprintnumbers,
showpacs,a4,twoside,twocolumn]{revtex4}

\usepackage{bm}
\usepackage{graphicx}
\usepackage{amsfonts}
\usepackage{amsmath}
\usepackage{amssymb}
\usepackage{amsbsy}
\usepackage{times}

\begin{document}

\title{Quantum critical scaling at a Bose-glass/superfluid transition: theory and experiment on a model quantum magnet}
\author{Rong Yu}
\affiliation{Department of Physics \& Astronomy, Rice University, Houston, TX 77005, USA}
\author{Corneliu F. Miclea} 
\affiliation{Condensed Matter and Magnet Science, Los Alamos National Lab, Los Alamos, NM 87545}
\affiliation{National Institute for Materials Physics,
077125 Bucharest-Magurele, Romania}
\author{Franziska Weickert} 
\affiliation{Condensed Matter and Magnet Science, Los Alamos National Lab, Los Alamos, NM 87545}
\author{Roman Movshovich}
\affiliation{Condensed Matter and Magnet Science, Los Alamos National Lab, Los Alamos, NM 87545}
\author{Armando Paduan-Filho} 
\affiliation{Instituto de Fisica, Universidade de S\~ao Paulo, 05315-970 S\~ao Paulo, Brasil}
\author{Vivien S. Zapf}
\affiliation{Condensed Matter and Magnet Science, Los Alamos National Lab, Los Alamos, NM 87545}
\author{Tommaso Roscilde}
\affiliation{Laboratoire de Physique, CNRS UMR 5672, Ecole Normale Sup\'erieure de Lyon, Universit\'e de Lyon, 46 All\'ee d'Italie, 
Lyon, F-69364, France}

% 03.75.Lm - Tunneling, Josephson effect, Bose-Einstein condensates in periodic
%            potentials, solitons, vortices and topological excitations
% 71.23.Ft - Quasicrystals
% 68.65.Cd - Superlattices
% 72.15.Rn - Localization effects (Anderson or weak localization)
\begin{abstract}
 In this paper we investigate the quantum phase transition  from magnetic Bose glass to magnetic Bose-Einstein
 condensation induced by a magnetic field in NiCl$_2\cdot$4SC(NH$_2$)$_2$ 
 (dichloro-tetrakis-thiourea-Nickel, or DTN), doped with Br (Br-DTN) or site diluted. Quantum Monte Carlo simulations for the 
 quantum phase transition of the model Hamiltonian for Br-DTN, as well as for site-diluted DTN, are consistent with conventional scaling
at the quantum critical point and with a critical exponent $z$ verifying the prediction $z=d$; moreover the correlation length exponent
 is found to be $\nu = 0.75(10)$, and the order parameter exponent to be $\beta = 0.95(10)$. We investigate the low-temperature thermodynamics at the quantum critical field of Br-DTN
 both numerically and experimentally, and extract the power-law behavior of the magnetization and of the specific heat. 
 Our results for the exponents of the power laws, as well as previous results for the scaling of the critical temperature to magnetic ordering
 with the applied field, are incompatible with the conventional crossover-scaling Ansatz proposed by Fisher \emph{et al.},  [Phys. Rev. B {\bf 40}, 546 (1989)], 
 but they can all be reconciled within a phenomenological Ansatz in the presence of a dangerously irrelevant operator.
\end{abstract}
\maketitle

\section{Introduction}

The investigation of interacting bosons in the presence of randomness
(the so-called \emph{dirty-boson} problem ) represents a long-standing subject in quantum condensed matter, and in particular a subject 
which remains far from being settled.
Indeed disorder has a tremendous impact on the physics of bosons, introducing a significant richness in their phase diagram both at zero and at finite temperature. 
Intense experimental investigations on a wide variety of different physical systems -- including $^4$He in porous media
\cite{Crowelletal97}, cold atoms in disordered optical potentials \cite{SanchezPalenciaL10}, and doped antiferromagnetic insulators \cite{Hongetal10,Yamadaetal11,Yuetal11,Huvonenetal12}, among others -- have been pursued in the attempt to unveil the universal aspects of such a phase diagram. A similarly intense activity has been conducted at the theoretical level \cite{Weichman08}, using a variety of methods ranging from field theory to quantum Monte Carlo.  
Yet, despite more than two decades of intense investigation, fundamental aspects of the physics of dirty bosons remain obscure. Among the most salient aspects under debate are the nature of the non-condensed and non-superfluid state induced by disorder \emph{and} interactions in a Bose fluid - namely the compressible Bose glass or the incompressible Mott glass; and the nature of the quantum phase transition (QPT) from a Bose glass to a strongly interacting superfluid Bose-Einstein condensate (BEC), induced by increasing the role of interactions while keeping the disorder strength fixed \cite{Fisheretal89}. In particular, the latter transition does not admit a well-controlled analytical treatment, its critical exponents are generally unknown, and the existing predictions \cite{Fisheretal89} are highly debated \cite{WeichmanM07}. 

In this paper we address the problem of the Bose-glass/superfluid QPT from the point of view of a model magnet, NiCl$_{2(1-x)}$Br$_{2x}\cdot$4SC(NH$_2$)$_2$  (Br-doped dichloro-tetrakis-thiourea-Nickel, or Br-DTN), whose magnetic Hamiltonian can be mapped exactly onto that of 3$d$ interacting bosons on a lattice with random hopping and random on-site interactions. A recent experiment \cite{Yuetal11} has demonstrated that this compound exhibits an extended magnetic Bose-glass phase in an applied magnetic field -- such a phase is characterized by a gapless and magnetically disordered low-temperature state. Increasing the field beyond a critical value drives the system through a QPT from Bose glass to magnetic BEC, characterized by spontaneous long-range antiferromagnetism in the plane transverse to the field. In particular the critical temperature to condensation is shown to scale with the applied field as $T_c \sim |h-h_c|^{\phi}$ with $\phi=1.1(1)$, where $h$ represents the dimensionless magnetic field and $h_c$ its critical value. 
A seemingly puzzling feature is that the only theoretical prediction for the 
$\phi$ exponent, stemming from Ref.~\onlinecite{Fisheretal89}, gives
$\phi \geq 2$, in open contradiction with the result of Ref.~\onlinecite{Yuetal11}. 
Using quantum Monte Carlo simulations of the model Hamiltonian for Br-DTN,  as well as of the Hamiltonian of DTN with a diluted magnetic lattice, 
we provide a detailed investigation of the quantum critical scaling at the Bose-glass/superfluid transition. In the case of Br-DTN we also investigate the low-temperature specific heat - both numerically and experimentally - and the magnetization - numerically - around the critical field. 
  Our theoretical data for the zero-temperature quantum critical point (QCP) support the picture of conventional scaling and the prediction \cite{Fisheretal89} of $z=d=3$ for the dynamical critical exponent. Nonetheless the results for the behavior of the system at finite temperature are seen to be globally incompatible with the predictions stemming from the scaling Ansatz of Ref.~\onlinecite{Fisheretal89} for the free energy, which assumes that the temperature is the only direction of instability of the QCP, while disorder and interactions can be eliminated from the scaling.
 We discuss possible violations of conventional scaling, due to the presence of a dangerously irrelevant operator, and find that a fully consistent picture can only be obtained 
when assuming that such an operator affects the finite-temperature behavior only, preserving conventional scaling at $T=0$. In the absence of a consistent theoretical treatment of the long-wavelength behavior of dirty bosons \cite{Weichman08}, our results offer a first, empirical determination of the scaling behavior around the
dirty-boson QCP.   

  The structure of the paper is as follows: Sec.~\ref{s.dbQCP} reviews the main aspects of the QPT of dirty bosons; Sec.~\ref{s.BrDTN} shortly illustrates how statically doped DTN realizes the physics of disordered lattice bosons;  Sec.~\ref{s.crossover} illustrates the crossover scaling Ansatz of Ref.~\onlinecite{Fisheretal89} and its predictions for the quantum critical behavior; Sec.~\ref{s.zd} discusses the numerical estimate of the dynamical critical exponent and of other critical exponents at the dirty-boson QCP; Sec.~\ref{s.QCtrajectory} presents our numerical as well as experimental results for the thermodynamics of Br-DTN at the quantum critical field; Sec.~\ref{s.ansatz} discusses how all of our results can be captured quantitatively by a generalized scaling Ansatz; conclusions are drawn in Sec.~\ref{s.conclusions}.

\section{Dirty-boson quantum critical point}
\label{s.dbQCP}

 In the absence of disorder, the physics of interacting bosons is generally well understood following a set of established paradigms. A diluted Bose gas of weakly interacting bosons undergoes condensation in a spatially extended state, and it is quantitatively described by Bogolyubov and Beliaev theory \cite{Griffin09}. The QPT from the vacuum state into the condensed state, driven by the chemical potential, 
  represents a Gaussian transition in dimensions $d>1$ with dynamical critical exponent $z=2$. \cite{Sachdev99} In $d=3$ the onset of the critical temperature to condensation around the QCP follows the power-law scaling $T_c \sim |\mu - \mu_c|^{\phi}$ (where $\mu_c$ is the critical chemical potential), with $\phi=2/3$ as predicted by mean-field theory \cite{Nikunietal00}. For bosons on a lattice, an instability to the formation of a Mott-insulating state occurs for sufficiently large interactions, and the QPT from a superfluid condensate to a Mott insulator away from commensurate filling admits the same description as the QCP of the diluted Bose gas, while it belongs to the universality class of the XY model in $d+1$ dimensions at commensurate filling. \cite{Fisheretal89}  
  
The introduction of disorder to the system is accompanied by the Anderson localization of the (low-energy) single-particle eigenstates; such states cannot host a Bose-Einstein condensate in the thermodynamic limit in the presence of any finite interaction.
As a consequence, the physics of the diluted Bose gas is completely altered by disorder, and a novel phase appears - the bosonic Anderson glass or Bose glass \cite{GiamarchiS88, Fisheretal89} - 
characterized by the redistribution of bosons on an extensive number of low-lying Anderson-localized states. This fact renders the theoretical description of the Bose glass quite involved, as conventional tools for the diluted Bose gas are no longer available. This is \emph{a fortiori} true when the interaction strength increases (the disorder strength being held fixed), inducing a QPT from the Bose glass into a strongly correlated condensed and superfluid state. This transition, as well as other transitions in disordered systems \cite{Parisi12}, represents an exceptionally hard problem in the theory of critical phenomena. Indeed the upper critical dimension $d_c$, above which mean-field theory becomes exact, is unknown for dirty bosons; and perturbative renormalization group treatments break down \cite{Fisheretal89}, given that they are based on the assumption of a finite $d_c$, while $d_c$ might as well be infinity for dirty bosons. This means that a quantitative understanding of the behavior of  interactions and disorder under renormalization is still lacking, as well as a rigorously motivated scaling form for the free energy. In the following we will offer evidence that the scaling Ansatz for the free energy proposed by Ref.~\onlinecite{Fisheretal89} is not appropriate. 

Another fundamental open question concerns the value of the dynamical critical exponent $z$.
The derivation of the identity $z=d$ offered in Ref.~\onlinecite{Fisheretal89} has been questioned in 
 Ref.~\onlinecite{WeichmanM07}, and some numerical calculations in $d=2$
 seem indeed to contradict it \cite{Priyadarsheeetal06, MeierW12}, although other 2$d$ studies are instead consistent with it
 \cite{ProkofevS04, Soyleretal11, Linetal11, Roscilde06, Yuetal08, Yuetal10}. For $d=3$ much less results are available, but $z=3$ is consistent with the existing numerical studies of the 3$d$ Bose-glass/superfluid transition \cite{HitchcockS06, Yuetal10}, and, as we will see, 
 also with the numerical results of this work (Sec.~\ref{s.zd}).
  
  The derivation of $z=d$ in Ref.~\onlinecite{Fisheretal89} is based on the observation that the compressibility
  $\kappa = \partial^2 f/\partial \mu^2$ (where $f$ is the free energy density and $\mu$ is the chemical potential)
  behaves as the imaginary-time stiffness of the system. An expansion of the \emph{singular} part of the free energy
  upon an infinitesimal twist leads to the prediction that $\kappa \sim g^{\nu(d-z)}$, where $g$ is the distance to the critical 
  point; if the compressibility is to remain finite across the transition, this would then imply that $d=z$. 
  This derivation has been questioned by Ref.~\onlinecite{WeichmanM07}, on the basis that the main response to a twist in 
  imaginary time is expected to come from the \emph{analytical} rather than from the singular part of the free energy -- this is due to the fact that, if the particle-hole symmetry is already broken (as it is across the generic Bose-glass/superfluid transition), a variation in the chemical potential does not alter the Hamiltonian symmetries in the system, and hence it should lead to negligible contributions in the singular part of the free energy density.  Therefore the finiteness of $\kappa$ across the transition should not have any obvious implication for the critical exponents.  
  
 It is generally hard to believe as an accident that many numerical results appear to be consistent with $z=d$ - although, as noticed in Ref.~\onlinecite{MeierW12}, only in a few cases an unbiased estimate of $z$ has been provided. Without the pretension of solving the issue of the value of $z$, in the following we provide a simple derivation of the finite-size scaling for the compressibility at zero temperature, based on its relationship with the particle-hole gap. Indicating with $E(N)$ the ground-state energy of the system having $N$ particles, we have that the particle-hole gap takes the form 
 \begin{equation}
\Delta_{\rm ph} = E(N+1) + E(N-1) - 2E(N). 
\end{equation}
For an hypercubic system of size $L^d$, taking the limit $L, N \gg 1$ one obtains 
\begin{equation}
\Delta_{\rm ph} \approx \frac{\partial^2 E}{\partial N^2} = \frac{1}{L^d} \frac{1}{\kappa}
\end{equation}
where we have used the fact that $\mu = \partial E/ \partial N$, and the definition of the compressibility $\kappa  = L^{-d} ~{\partial N}/{\partial \mu}$.
At the QCP, the finite-size particle-hole gap must vanish as $\Delta_{\rm ph} \sim L^{-z}$, yielding the result $\kappa \sim L^{-(d-z)}$. 
Hence a finite compressibility in the thermodynamic limit at the QCP would require that $z=d$. 
The above derivation is based on the only assumption that  conventional scaling applies (giving $\Delta_{\rm ph} \sim L^{-z}$), namely that the system is below or at the upper critical dimension. 
The scaling of $\kappa$ is indeed observed at the commensurate-incommensurate transition of the Bose-Hubbard model with $z=2$ both in $d=1$ (where the compressibility diverges \cite{Giamarchi04}) and in $d=2$ (where the compressibility jumps \cite{AletS04}, up to logarithmic corrections). It is also verified at the Mott-insulator/superfluid transition at commensurability (with $z=1$), \emph{e.g.} in $d=1$ (compressibility jump at the Kosterlitz-Thouless transition \cite{Giamarchi04}) and in $d=2, 3$ (where the compressibility vanishes continuously \cite{Wangetal06, Nohadanietal05}). 

\section{Dirty bosons and the physics of doped DTN}
\label{s.BrDTN}

Recently a new experimental playground has become available for the study of the physics of dirty bosons: doped magnetic insulators. In such systems a Bose fluid of magnetic quasiparticles (corresponding to modes carrying a spin $m_s=1$) can be controlled by the application of a magnetic field, and can be made to condense at sufficiently low temperature when the field exceeds a critical value \cite{Giamarchietal08}. The condensed phase corresponds to long-range (anti)ferromagnetism in the plane transverse to the field. Static doping of the magnetic insulator leads to disorder in the magnetic Hamiltonian, and allows therefore to investigate a field-induced Bose-glass/superfluid transition, which has been discussed theoretically \cite{RoscildeH05, Nohadanietal05-2, Roscilde06, Yuetal11} as well as recently probed in experiments \cite{Hongetal10, Yuetal11, Yamadaetal11, ZheludevH11, Yamadaetal11-2, Huvonenetal12}.
 
 Here we focus our attention on Br-doped DTN, which has been recently investigated in Ref.~\onlinecite{Yuetal11}.
 A minimal model Hamiltonian for the magnetic behavior of Br-DTN is given by 
\begin{eqnarray}
{\cal H}_{\rm Br-DTN} &=& \sum_{\langle ij \rangle_{c}}  
J_{c,\langle ij \rangle} ~{\bm S}_{i}\cdot{\bm S}_{j} ~~+ 
J_{ab} \sum_{\langle lm \rangle_{ab}} 
{\bm S}_{l}\cdot{\bm S}_{m} \nonumber \\
&+& \sum_{i} D_i (S^z_{i})^2
- g\mu_B H \sum_{i}  S^z_{i}.
\label{e.Ham}
\end{eqnarray}
Here $S_i^{\alpha}$ ($\alpha=x,y,z$) are $S=1$ spin operators, coupled on a cubic lattice with antiferromagnetic  interactions $J_c$ along the $c$ axis and $J_{ab}$ in the $ab$ plane; $D$ is a strong single-ion anisotropy. In pure DTN $J_{c}=2.2$ K, $J_{ab}=0.18$ K, $D=8.9$ K and $g=2.26$. 
\cite{Zvyaginetal07}
The effect of Br doping on the magnetic behavior can be quantitatively captured by promoting the $J_c$ couplings and the single-ion anisotropies to local, bond-dependent ($J_{c,\langle ij \rangle}$) and site-dependent ($D_i$) quantities, related to the appearance of a Br dopant on the Cl-Cl bond bridging the antiferromagnetic interactions between Ni ions along the $c$ axis. In particular a concentration $x$ of Br dopants leads to a fraction $2x$ of doped Cl-Br bonds (neglecting the case of rare Br-Br bonds for $x \ll 1$); Br doping is modeled as increasing the $J_c$ bond strength to $J_c'=2.3 J_c$, and reducing the single anisotropy $D$ to $D'=D/2$ on one of the two ions connected by the bond. In the rest of this paper we will focus on the doping value $x = 0.075$. 

A Holstein-Primakoff transformation maps the magnetic Hamiltonian
 onto a lattice boson Hamiltonian \cite{Yuetal11}
\begin{eqnarray}
{\cal H} &=&
- \sum_{\langle ij \rangle} J_{ij}
\left[ \sqrt{1-\frac{n_i}{2}}~ b_i b_j^{\dagger} ~
\sqrt{1-\frac{n_j}{2}}  + {\rm h.c.} \right] \nonumber \\
&&
+ \sum_{\langle ij \rangle} J_{ij} (n_i-1)(n_j-1)
+ \sum_i D_i (n_i - 1)^2 \nonumber \\
&& - g\mu_B H \sum_{i}  n_i + {\rm const.} ~~~~~~
\label{e.Bosehamilton}
\end{eqnarray} 
where $b_i, b^{\dagger}_i$ are bosonic operators with maximum occupation $n_i = 2$;  here $J_{ij} = J_{c, \langle ij \rangle}$, $J_{ab}$ depending on the orientation of the bond. The single-ion anisotropy plays the role of an on-site repulsion term for the bosons (which penalizes doubly occupied as well as empty sites, due to the spin-inversion symmetry of the Hamiltonian); the antiferromagnetic couplings play the role of a hopping term as well as of nearest-neighbor repulsion; and the role of the chemical potential is taken by the applied magnetic field. 
In particular the pure system in zero field is in a Mott-insulating phase with $n=1$ particle per site (corresponding to zero magnetization on each site), and 
it exhibits a Mott-insulator/superfluid QPT at a critical field $h_{c1} = g \mu_B H/J_c \approx 1.65$; a second transition from superfluid to $n=2$ Mott insulator is reached at the upper critical field $h_{c2}=8.69$. In the presence of Br doping $h_{c1}$ shifts to a lower value $h_{c1} = 0.828(3)$ and the QPT connects  a Bose-glass to a superfluid state \cite{Yuetal11}; $h_{c2}$ is instead weakly affected by Br doping, but the corresponding transition also changes to a Bose-glass-to-superfluid one. 

 To probe the universal character of some of our findings, we consider as well the case of site dilution of the magnetic lattice of DTN, leading to the following model Hamiltonian 
 \begin{eqnarray}
{\cal H}_{\rm sd-DTN} &=& J_c  \sum_{\langle ij \rangle_{c}}  ~ \epsilon_i \epsilon_j {\bm S}_{i}\cdot{\bm S}_{j} ~~+ 
J_{ab} \sum_{\langle lm \rangle_{ab}} \epsilon_i \epsilon_j 
{\bm S}_{l}\cdot{\bm S}_{m} \nonumber \\
&+& D \sum_{i} \epsilon_i (S^z_{i})^2
- g\mu_B H \sum_{i} \epsilon_i S^z_{i}.
\label{e.Hamsd}
\end{eqnarray}
The Hamiltonian parameters are now set to be homogeneous and equal to those of pure DTN, but the random variables $\epsilon_i$ have been introduced, taking the value 0 with probability $x$ and 1 with probability $1-x$. Here $x$ is the lattice dilution. The phase diagram of the model Eq.~\ref{e.Hamsd} has been extensively studied in Ref.~\onlinecite{Yuetal10-2} for a dilution $x=0.15$, and it exhibits two superfluid/Bose-glass transitions at fields $h_{c1} \approx 1.94(1)$  and 
$h_{c2} \approx 8.52(2)$. Site dilution of the lattice of magnetic Ni ions in DTN can be achieved by Cd doping \cite{Yinetal11}. 
Similarly to the case of Br-DTN, a spin-boson transformation maps the spin Hamiltonian Eq.~\eqref{e.Hamsd} to that of a lattice gas of strongly interacting bosons hopping on a spatially anisotropic cubic lattice with site dilution.

In the following we will mainly focus on the vicinity of the critical points $h_{c1}$ and $h_{c2}$. As sketched in Fig.~\ref{f.QCcartoon}),
for $d=3$ the $T=0$ QCPs are the origin of a line of critical temperatures $T_c(g)$ ($g = |h-h_{c1(2)}|$) for the transition to a superfluid condensed phase. 
The critical temperature is follow the power-law dependence $T_c(g) \sim g^\phi$. In the pure system, $\phi=2/3$, stemming from the theory of the diluted Bose gas \cite{Nikunietal00}; this result has also been established experimentally with high accuracy for DTN \cite{Yinetal08}. 
 In the disordered case, both experiment and theory find $\phi = 1.1(1)$ for Br-DTN; this result is also confirmed theoretically in site-diluted DTN. \cite{Yuetal10-2}
 It is remarkable to observe that other recent experiments on different compounds, expected to exhibit a field-induced Bose-glass/superfluid transition, also report values of $\phi$ which are compatible with what found in DTN. 
 In particular a $\phi \approx 1$ is compatible with the results on K-doped TlCuCl$_3$, \cite{Yamadaetal11}, as commented in Ref.~\onlinecite{ZheludevH11}
 (although Ref.~\onlinecite{Yamadaetal11} claims that an exponent $\phi \sim 2$ can be extracted from a fit to the experimental data, see also
 Ref.~\onlinecite{Yamadaetal11-2}). Values of $\phi$ compatible with those of doped DTN are also found in a recent experiment on piperazinium-Cu$_2$(Cl$_{1-x}$Br$_x$)$_6$. \cite{Huvonenetal12}. 
 
 This mounting theoretical and experimental evidence in favor of $\phi\approx 1$ for the $T_c$ line stemming from the dirty-boson QCP clashes with the only prediction existing in the literature, namely that of Ref.~\onlinecite{Fisheretal89}, giving $\phi = \nu z \geq 2$ if $z=d$. In the following section we will shortly review the scaling Ansatz on the free energy of dirty bosons from which the above prediction stems, as well as its further implications for measurable properties.

\section{One-argument crossover scaling Ansatz}
\label{s.crossover}

\begin{figure}[h]
\begin{center}
\includegraphics[
%bbllx=60pt,bblly=50pt,bburx=510pt,bbury=450pt,%
    width=60mm,angle=0]{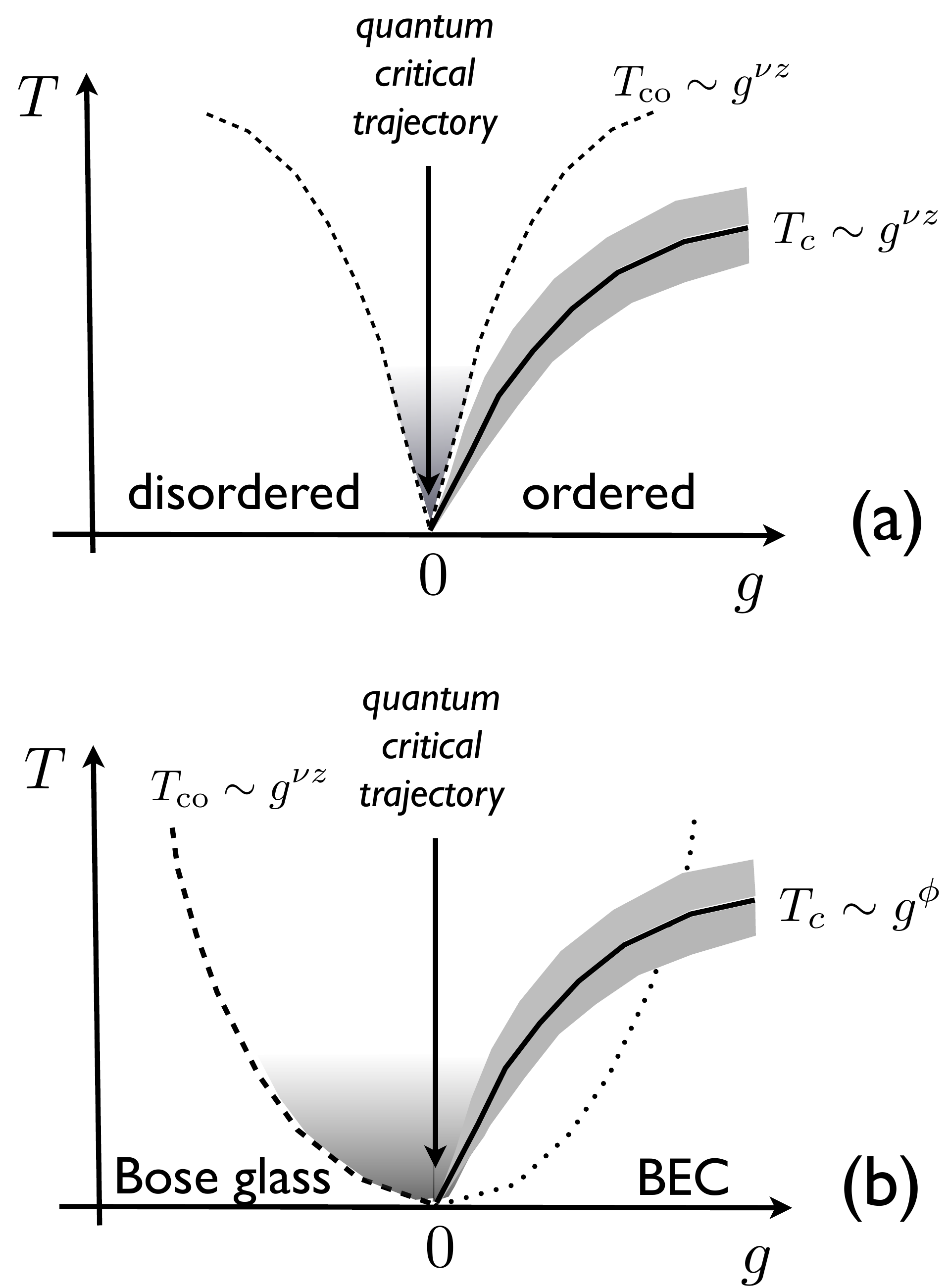}
\end{center}
\caption{(a) At a QPT, the quantum critical behavior at finite temperature manifests itself within a "fan" loosely bounded from below by crossover lines
$T_{\rm co}\sim g^{\nu z}$. The line of thermal transitions $T_c\sim g^{\phi}$, and the surrounding Ginzburg region (gray-shaded area) of classical critical behavior, can be 
well decoupled from the quantum critical fan if $\phi = \nu z$. (b) \emph{Conjectured} picture for the dirty-boson QPT: for dirty bosons $\phi < \nu z$, so that 
the hierarchy between $T_c(g)$ line and the putative crossover temperature $T_{\rm co}$ to quantum-critical behavior is actually inverted. This would imply that the Ginzburg region and the quantum critical region touch each other.}
\label{f.QCcartoon}
\end{figure}

 The onset of a finite-temperature transition at $T_c(h)$ starting 
 from the QCPs at $h_{c1}$, $h_{c2}$ is a complex problem
 in the theory of critical phenomena, because the thermal transition and the quantum transition
 are completely different in nature. A comprehensive treatment of the two can be attempted via 
 the crossover scaling theory of phase transitions \cite{Cardybook}, which can account for 
 the crossover between the quantum critical behavior at low and zero temperature
 and in the vicinity of the QCP, and the classical critical behavior close to the line of 
 critical temperatures $T_c(h)$ (see Fig.~\ref{f.QCcartoon} for a sketch).  
 For a system of bosons in a disordered environment Ref.~\onlinecite{Fisheretal89} propose 
  the simplest possible \emph{(one-argument)} crossover scaling Ansatz for the singular part of 
 the free-energy density, namely
 \begin{equation}
 f_s(T,g) \sim |g|^{2-\alpha} F\left( \frac{T}{|g|^{\nu z}} \right) = T^{\frac{2-\alpha}{\nu z}} G\left( \frac{T}{|g|^{\nu z}} \right)
 \label{e.onepscaling}
 \end{equation}
where $g = h-h_{c1(2)}$ and $\nu$, $z$ and $\alpha$ are the critical exponent of the QCP. Here $T$ is in units
of $J_c/k_B$. 
Such a scaling Ansatz can be obtained via a phenomenological assumption for
the behavior of the free energy under renormalization group (RG) \cite{Cardybook}.
Such a behavior is valid when the QCP is a fixed point under RG which only admits 
the temperature $T$ and the control parameter $g$ as directions of instability.
All other directions in parameter space (the strength of disorder, the strength of 
interactions, etc.) are supposed to be stable directions, and the associated parameters can be
eliminated from the scaling Ansatz.  

In order for the system to exhibit a finite-temperature transition at $T_c$, 
the function $F(y)$ must exhibit a singularity at a given value $y_c$ of its argument, 
so that one expects the critical temperature to follow the scaling law
$T_c = y_c |g|^{\phi}$  with scaling exponent $\phi=\nu z$. This means that, according to the scaling
Ansatz Eq.~\eqref{e.onepscaling}, the onset of the critical temperature is 
completely governed by the critical exponents $\nu$ and $z$ of the QPT.  
Assuming that $z=d$ for the dirty-boson QCP,\cite{Fisheretal89} and given
that $\nu\geq 2/d$ according
to the Harris criterion \cite{Chayesetal86}, one obtains a scaling exponent $\phi = \nu z \geq 2$, as already mentioned previously. 

 Further consequences of the scaling Ansatz Eq.~\ref{e.onepscaling} involve the thermodynamic behavior 
 of the system along the quantum critical trajectory, $T\to 0$ for $g =0$. In fact one
 has that the specific heat $C$ scales as
 \begin{equation}
 C \sim T \frac{\partial^2 f_s}{\partial T^2}{\Big |}_{g=0} \sim T^{x_C}
 \label{e.Cscaling}
  \end{equation}
  with 
  \begin{equation}
  x_C = \frac{2-\alpha}{\nu z}-1
  \label{e.xc}
  \end{equation}
while the uniform (field-induced) magnetization scales as 
 \begin{equation}
 m(T) - m(0) \sim \frac{\partial f_s}{\partial g}{\Big |}_{g=0} \sim T^{x_m}
  \label{e.mscaling}
  \end{equation}
  with
  \begin{equation}
  x_m = \frac{1-\alpha}{\nu z}~.
  \label{e.xm} 
 \end{equation}
 The scaling Ansatz Eq.~\eqref{e.onepscaling} is completely consistent with
 the validity of conventional scaling at the QCP, and in particular of hyperscaling, $\nu(d+z) = 2-\alpha$, 
 so that one further obtains  $x_C = (d+z)/z - 1 = 1$ and $x_m = 2 - 1/(\nu z) \geq 3/2$. 

The prediction $\phi \geq 2$ is based on the assumption that $z=d$, which, as discussed in Sec.~\ref{s.dbQCP}, is still a matter of debate. 
 Hence one could argue that the prediction of Ref.~\onlinecite{Fisheretal89} can be reconciled with the observations that $\phi\approx 1$ in disordered magnets by assuming that 
$z=d$ is indeed violated. In the next section we will provide an accurate study of the quantum critical scaling at the $T=0$ QCP for both Br-doped DTN and site-diluted DTN, testing the validity of the $z=d$ prediction.

 \section{Finite-size scaling for the Br-DTN Hamiltonian}
 \label{s.zd}

 In this section we provide a finite-size scaling analysis of quantum Monte Carlo results for the dirty-boson QCP exhibited by the models for doped DTN (Br-doped or site-diluted). Our quantum Monte Carlo results have been obtained making use of the Stochastic Series Expansion (SSE) approach \cite{SSE} on $L^3$ lattices with  $L=$ 12, 14, 16 and 18. All results are averaged over $\sim 300$ disorder realizations.   
 The quantities of interest here are: 1) the correlation length along the $c$ axis and in $ab$ plane, obtained via the second-moment estimator
 \begin{equation}
 \xi_{p} = \frac{L}{2\pi} \sqrt{\frac{S^{\perp}(\bm Q)}{S^{\perp}(\bm Q+\bm q)} - 1};
 \end{equation}
 where
 \begin{equation}
 S^{\perp}({\bm k}) = L^{-3} \sum_{ij} e^{i{\bm k}\cdot ({\bm r}_i- {\bm r}_j)} \langle S_i^x S_j^x + S_i^y S_j^y \rangle 
 \end{equation}
 is the structure factor for the spin components transverse to the applied field, $\langle ... \rangle$ denotes the thermal and disorder average,
 $\bm Q = (\pi, \pi, \pi)$ is the ordering vector, and ${\bm q} = (2\pi/L) \hat{c}$ for $p=c$ and 
 ${\bm q} = (2\pi/L) \hat{a}$ ~(or $(2\pi/L)\hat{b}$) for $p=ab$; 2) the spin stiffness 
 \begin{equation}
 \rho_s = \frac{k_B T}{4J_c L} \langle |\bm W|^2\rangle 
 \end{equation}
 where ${\bm W} = (W_a, W_b, W_c)$ is the winding number of bosonic worldlines along the three spatial directions
 \cite{PollockC87}; 3) the squared order parameter, defined as $m_s^2 = S^{\perp}({\bm Q})/L^3$. 
 
 \begin{figure}[h]
\begin{center}
\includegraphics[
%bbllx=60pt,bblly=50pt,bburx=510pt,bbury=450pt,%
    width=90mm,angle=0]{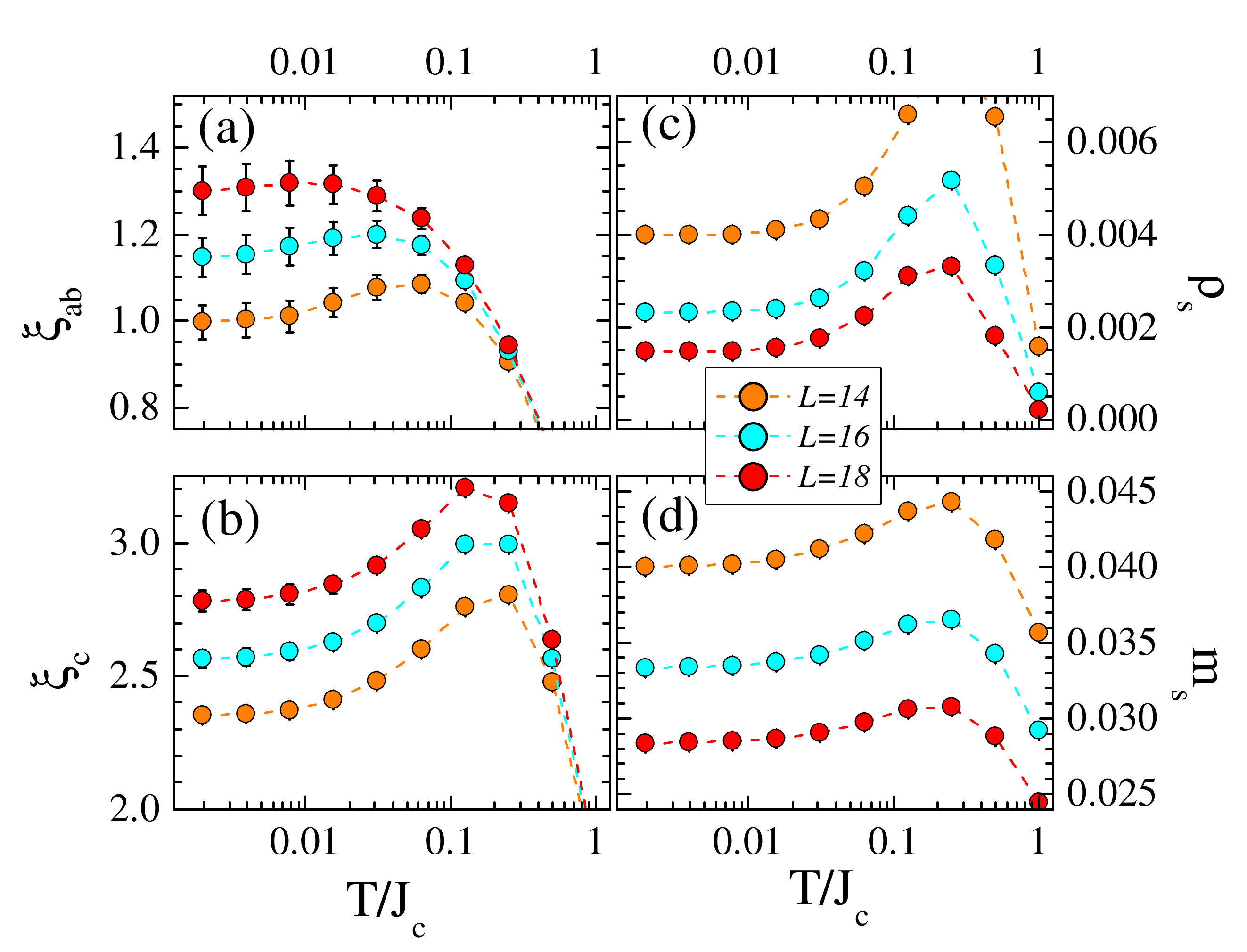}
\end{center}
\caption{$\beta$-doubling plots for the correlation length, spin stiffness and order parameter of the Br-DTN Hamiltonian, $h = 0.83$.}
\label{f.betadouble}
\end{figure}    
 
 We have applied a $\beta$-doubling procedure \cite{Sandvik02}, particularly fit for SSE, which consists in decreasing exponentially the temperature until convergence of the above quantities is reached towards their $T=0$ value. Fig.~\ref{f.betadouble} shows that, for the system sizes under investigation, this approach leads successfully to the physical $T=0$ value of the quantities of interest for temperatures $T \approx 2*10^{-3} J_c$. In particular Fig.~\ref{f.betadouble} shows that convergence is achieved even in the most delicate case $h = 0.83$ for Br-doped DTN, corresponding to a field very close to the QCP. 

 If conventional scaling applies to the system - namely, if 3$d$ dirty bosons are below their upper critical dimension - then one expects the following finite-size scaling behavior for the above quantities: 
 \begin{eqnarray}
 \xi_p(g,L) &=& L ~F_{\xi_p}\left(|g|L^{1/\nu}\right) \label{e.xi} \\
 \rho_s(g,L) &=& L^{d+z-2} ~F_{\rho_s}\left(|g|L^{1/\nu}\right) \label{e.rhos} \\
 m_s^2(g,L) &=& L^{2\beta/\nu} ~F_{m_s}\left(|g|L^{1/\nu}\right) \label{e.ms} 
 \end{eqnarray}
 
 For finite-temperature data, one should also include the temperature dependence of the scaling functions as $F_O(|g|L^{1/\nu}, \beta L^{-z})$ for the observable $O$, and hence one would need a prior knowledge of the $z$ exponent to be able to obtain the universal functions from collapse of the universal data. On the other hand we can omit the temperature argument of the scaling functions, having systematically eliminated thermal effects from our data; this will allow us to give an unbiased estimate of the critical exponents.  
 
 \begin{figure}[h]
\begin{center}
\includegraphics[
%bbllx=60pt,bblly=50pt,bburx=510pt,bbury=450pt,%
    width=90mm,angle=0]{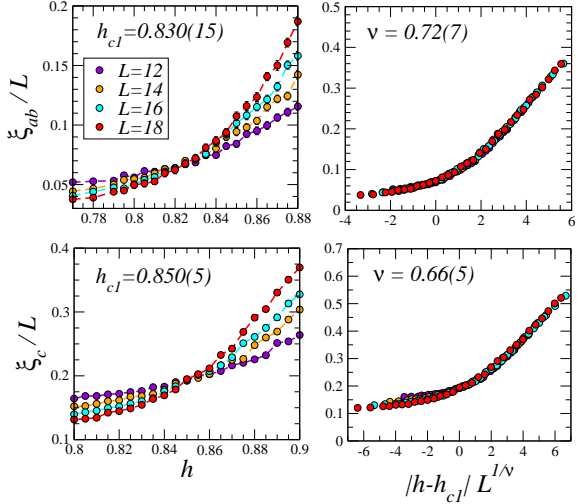}
\end{center}
\caption{Scaling plots of the QMC data for the correlation length of the Br-DTN model.}
\label{f.xi-QCscaling}
\end{figure}

 \subsection{Br-doped DTN}
 
 We can make use of Eq.~\eqref{e.xi} to locate the critical field via the intersection of the $\xi_p/L$ curves plotted as a function of $h$. This is shown in Fig.~\ref{f.xi-QCscaling}. The crossing point for the $\xi_{ab}/L$ curves is located at $h_{c1} = 0.830(15)$, with a fairly large error bar due to the shallow form of the curves around the critical field. 
 In particular this value is in very good agreement with the value $h_{c1} = 0.828(3)$ obtained by extrapolation of the $T_c(h)$ curve to zero temperature
 \cite{Yuetal11}. On the other hand, we observe that the crossing point for the $\xi_c/L$ curves is at a slightly larger value, $h_{c1} = 0.850(5)$.  
 We attribute this discrepancy to the spatial anisotropy present in the Br-DTN model in terms of disorder effects, given that bond randomness only affects the $J_c$ couplings and not the $J_{ab}$ ones. This asymmetry might hinder the development of correlations along the $c$ axis, and hence alter the finite-size estimates of the critical field. Nonetheless a very good collapse is obtained in a scaling plot as a function of $gL^{1/\nu}$ with $\nu = 0.72(7)$ for the $\xi_{ab}$ curves and 
 $\nu=0.66(5)$ for the $\xi_c$ curves (only for positive $g$ for the latter, again showing an asymmetry between the two spatial directions). 
 
 \begin{figure}[h]
\begin{center}
\includegraphics[
%bbllx=60pt,bblly=50pt,bburx=510pt,bbury=450pt,%
    width=70mm,angle=0]{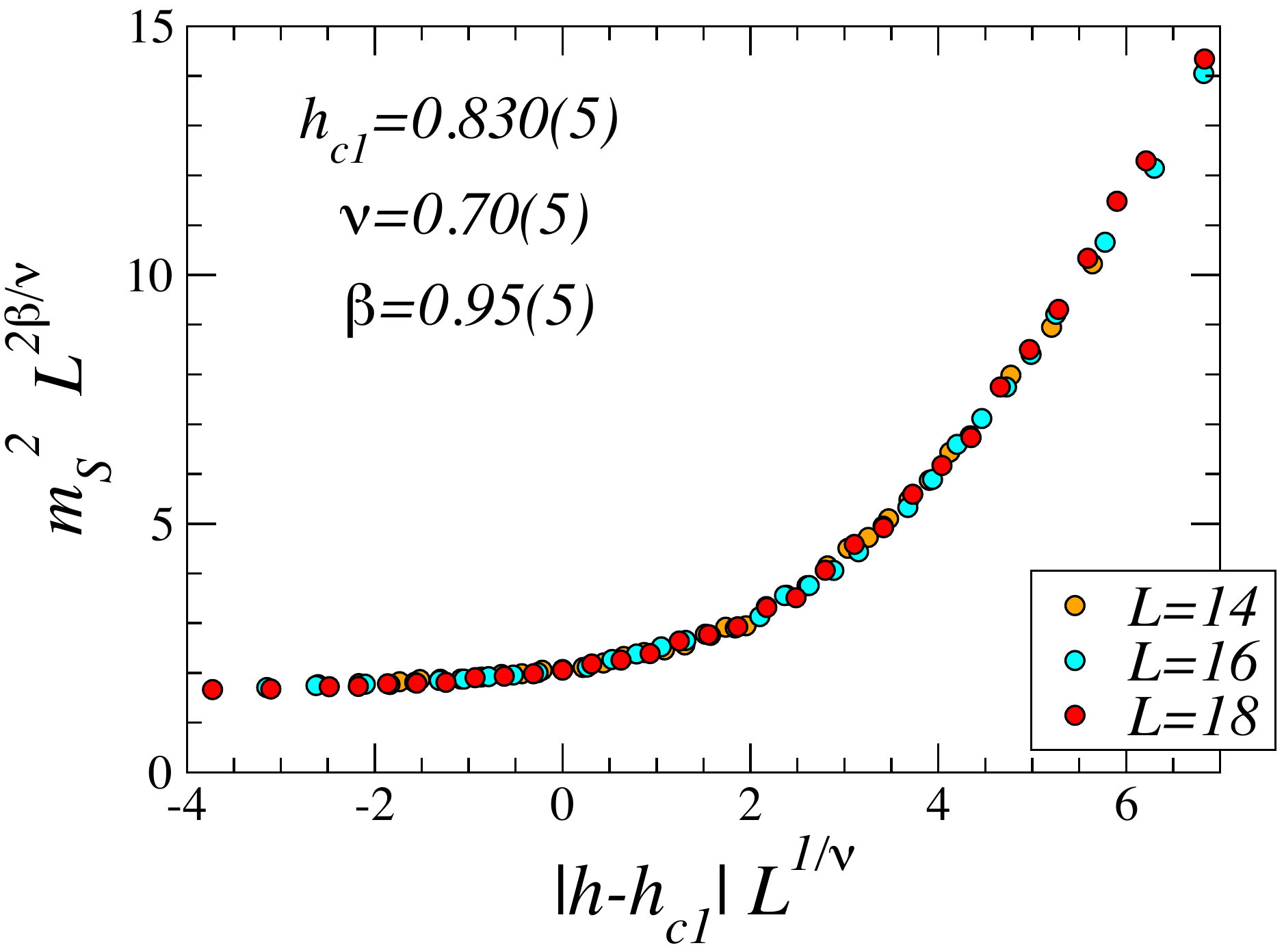}
\end{center}
\caption{Scaling plots of the QMC data for the squared order parameter of the Br-DTN model.}
\label{f.ms-QCscaling}
\end{figure}

 To better refine our estimate of $h_{c1}$ we consider the scaling of the order parameter, depicted in Fig.~\ref{f.ms-QCscaling}. We observe an excellent collapse for $h_{c1} = 0.830(5)$, $\nu = 0.70(5)$ and $\beta=0.95(5)$. On the other hand, we cannot obtain such a good collapse both for negative and positive $g$ values by using the larger $h_{c1}$ estimate given by the scaling of $\xi_{c}$. Hence in the following we will make use of the $h_{c1}\approx 0.83$ estimate. 
 \begin{figure}[h]
\begin{center}
\includegraphics[
%bbllx=60pt,bblly=50pt,bburx=510pt,bbury=450pt,%
    width=90mm,angle=0]{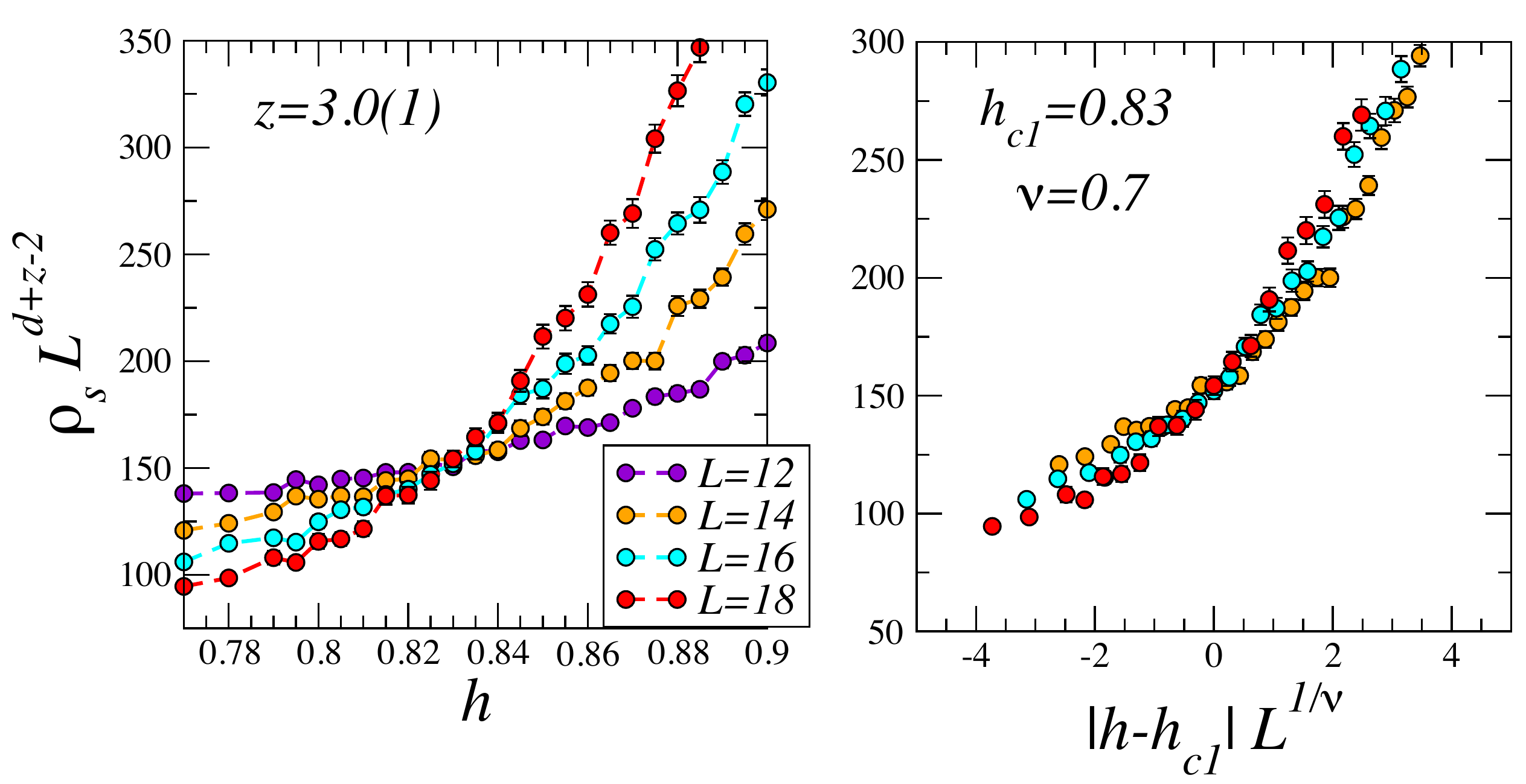}
\end{center}
\caption{Scaling plots of the QMC data for the spin stiffness of the Br-DTN model.}
\label{f.rhos-QCscaling}
\end{figure}   
 We conclude this analysis by examining the behavior of $\rho_s$. Imposing that the $\rho_s L^{d+z-2}$ curves cross at the estimated $h_{c1}$, we obtain that $z=3.0(1)$, fully consistent with $z=d$ (see Fig.~\ref{f.rhos-QCscaling}). A full scaling plot of $\rho_s$ with $\nu \approx 0.7$ does not give a very satisfactory collapse far from the critical point. We attribute this lack of scaling to the asymmetry between the $ab$ plane and the $c$ axis: indeed $\rho_s$ contains the winding number fluctuations along all spatial directions, whose scaling properties for the sizes we have examined might be different. We postpone to future work a detailed analysis of the winding number fluctuations along specific spatial directions. Another possibility is that corrections to scaling must be taken into account in the behavior of $\rho_s$ for the system sizes considered here (although they did not seem necessary to account for the scaling of the other observables).  
 \begin{figure}[h]
\begin{center}
\includegraphics[
%bbllx=60pt,bblly=50pt,bburx=510pt,bbury=450pt,%
    width=70mm,angle=0]{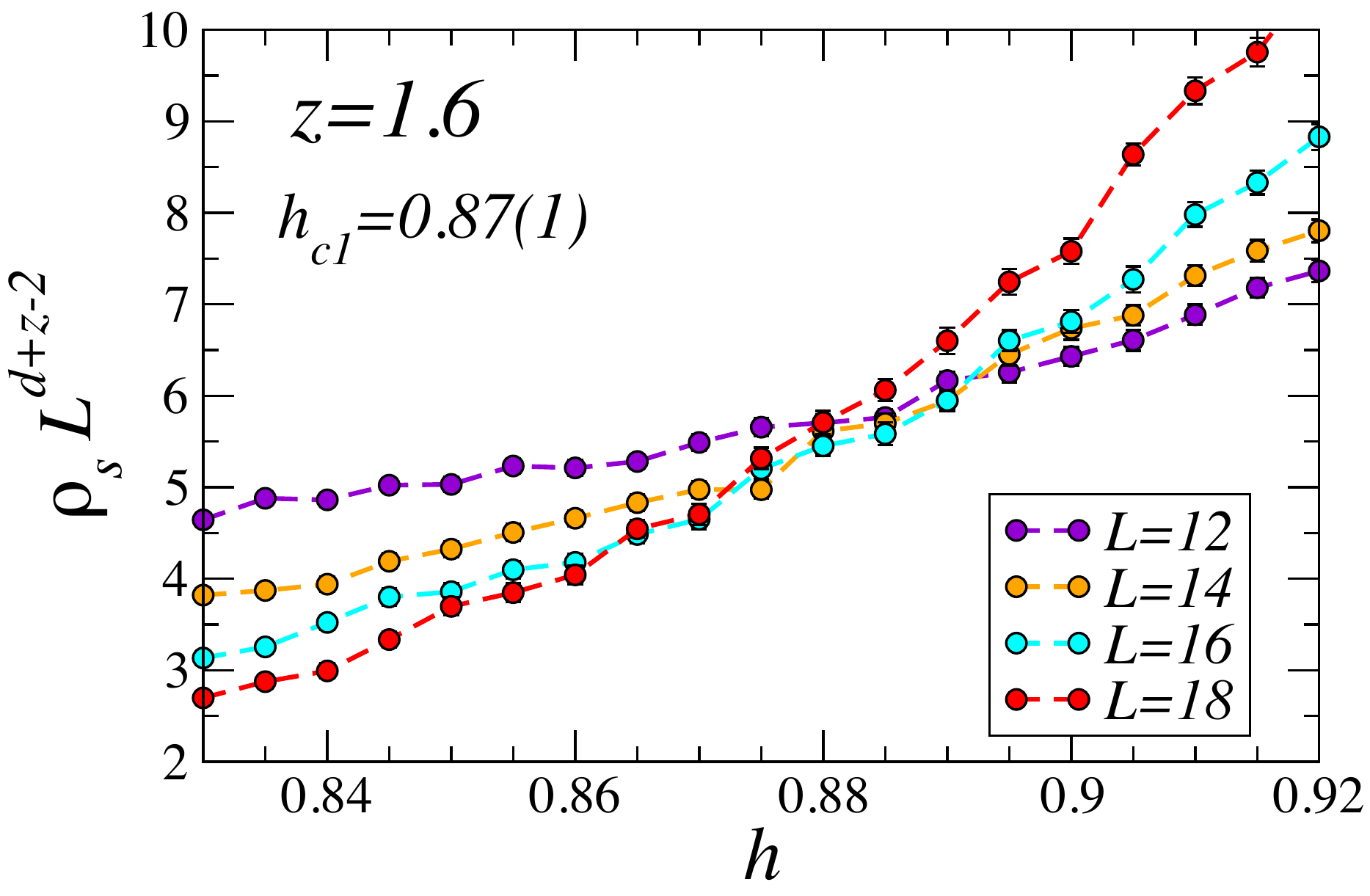}
\end{center}
\caption{Scaling plot of the QMC data for the spin stiffness of the Br-DTN model using $z=1.6$. The scaled curves do not appear to cross at the same field.}
\label{f.wrhos-QCscaling}
\end{figure}     

  Relying on the conclusions of Sec.~\ref{s.crossover}, one would argue that $\nu z = \phi$ would rather imply that $z = \phi/\nu \approx 1.6$ (using $\phi = 1.1$ and $\nu = 0.7$). On the other hand, when using this value for $z$ we obtain that the $\rho_s(L) L^{d+z-2}$ curves
 do not cross at one and the same critical field -- see Fig.~\ref{f.wrhos-QCscaling}. Willing to identify an estimate of the critical 
 field from the crossing of the two largest system sizes ($L=16$ and 18) we obtain
 $h_{c1} = 0.87(1)$, clearly inconsistent with the previous estimates. Hence the equality $\phi = \nu z$ is definitely incompatible with our results. 
 \begin{figure*}[ht!]
\begin{center}    
\includegraphics[
%bbllx=60pt,bblly=50pt,bburx=510pt,bbury=450pt,%
    width=85mm,angle=0]{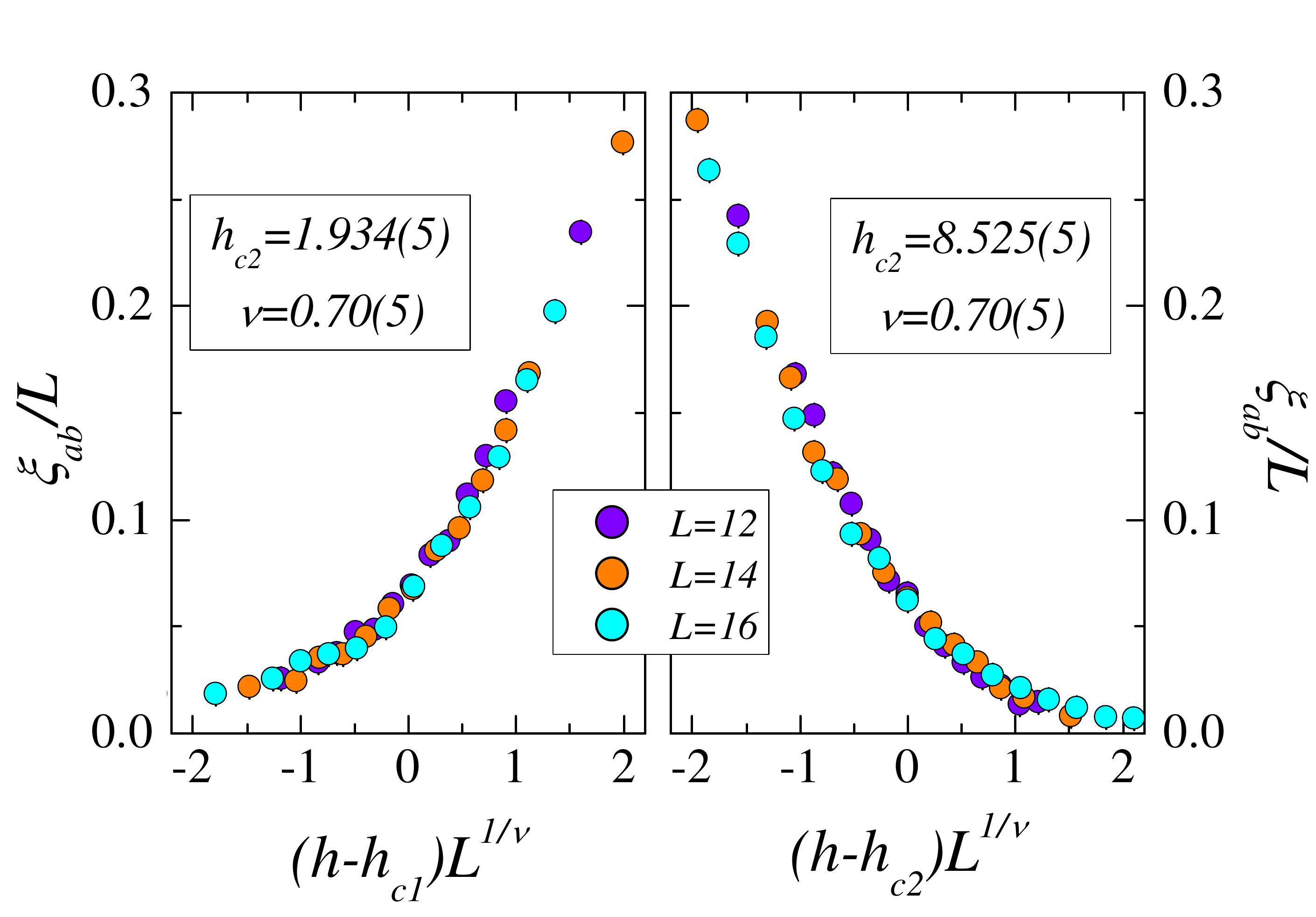} ~~~~~~~~~
\includegraphics[
%bbllx=60pt,bblly=50pt,bburx=510pt,bbury=450pt,%
    width=85mm,angle=0]{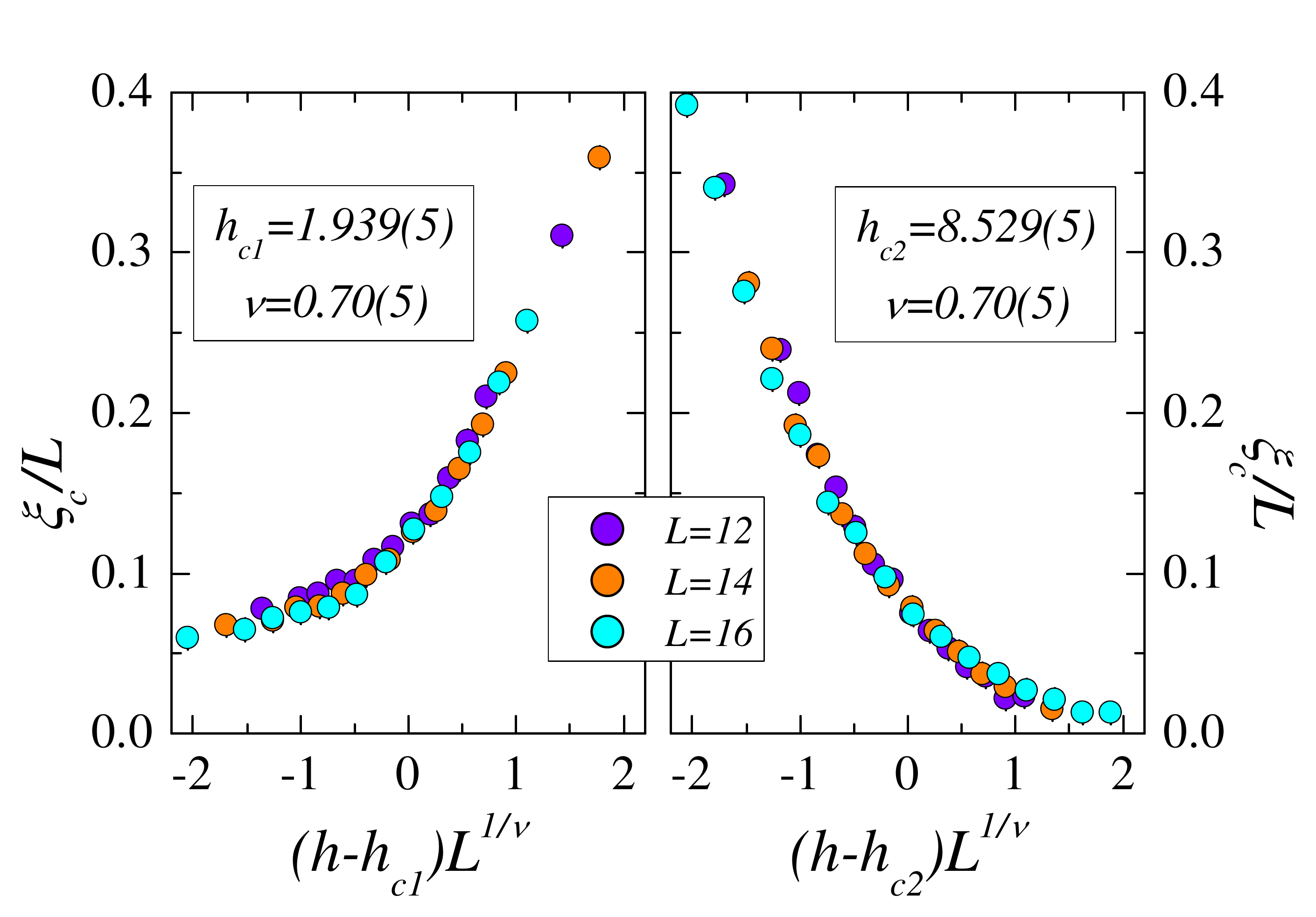}
\includegraphics[
%bbllx=60pt,bblly=50pt,bburx=510pt,bbury=450pt,%
    width=85mm,angle=0]{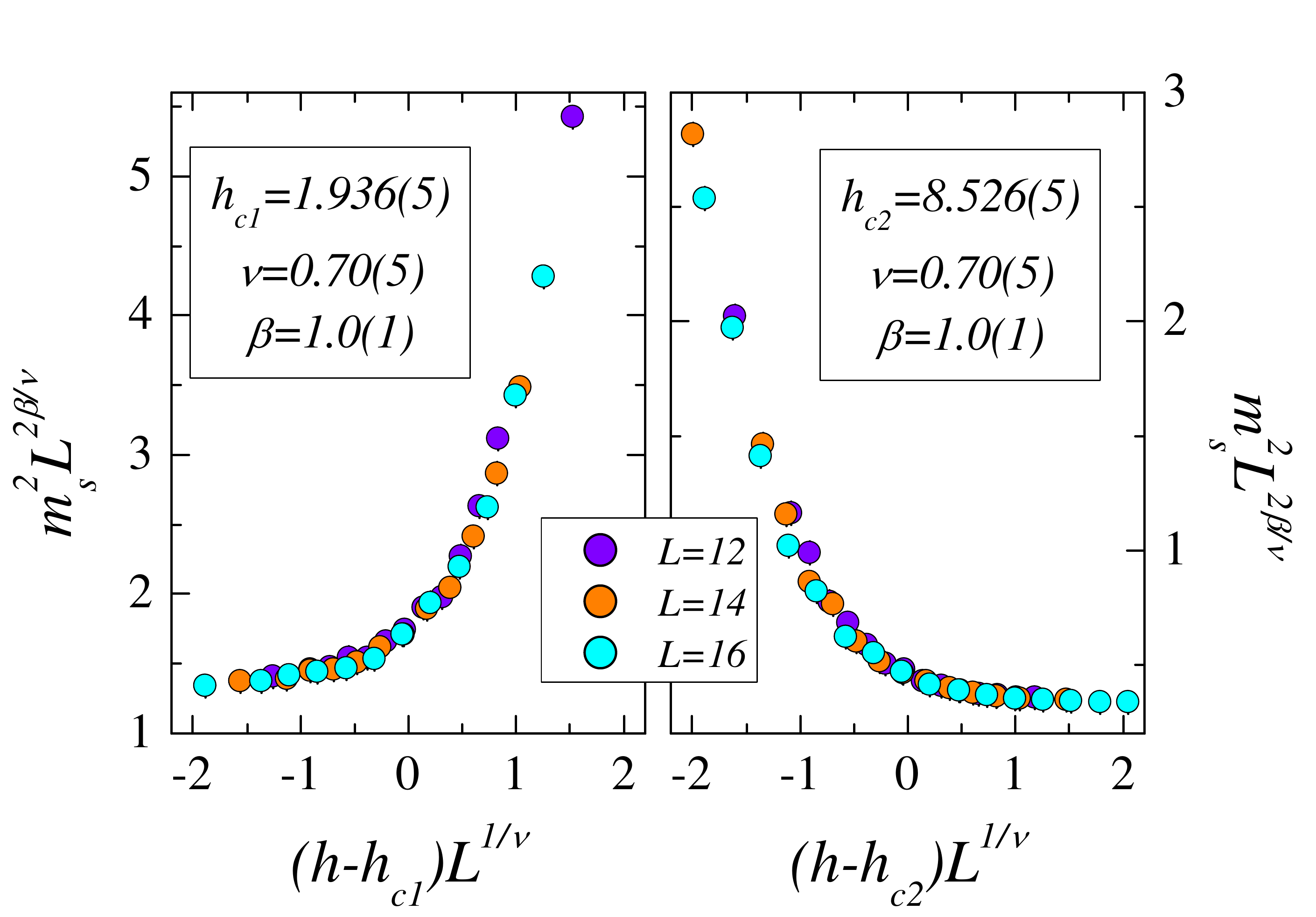}    ~~~~~~~~~
\includegraphics[
%bbllx=60pt,bblly=50pt,bburx=510pt,bbury=450pt,%
    width=85mm,angle=0]{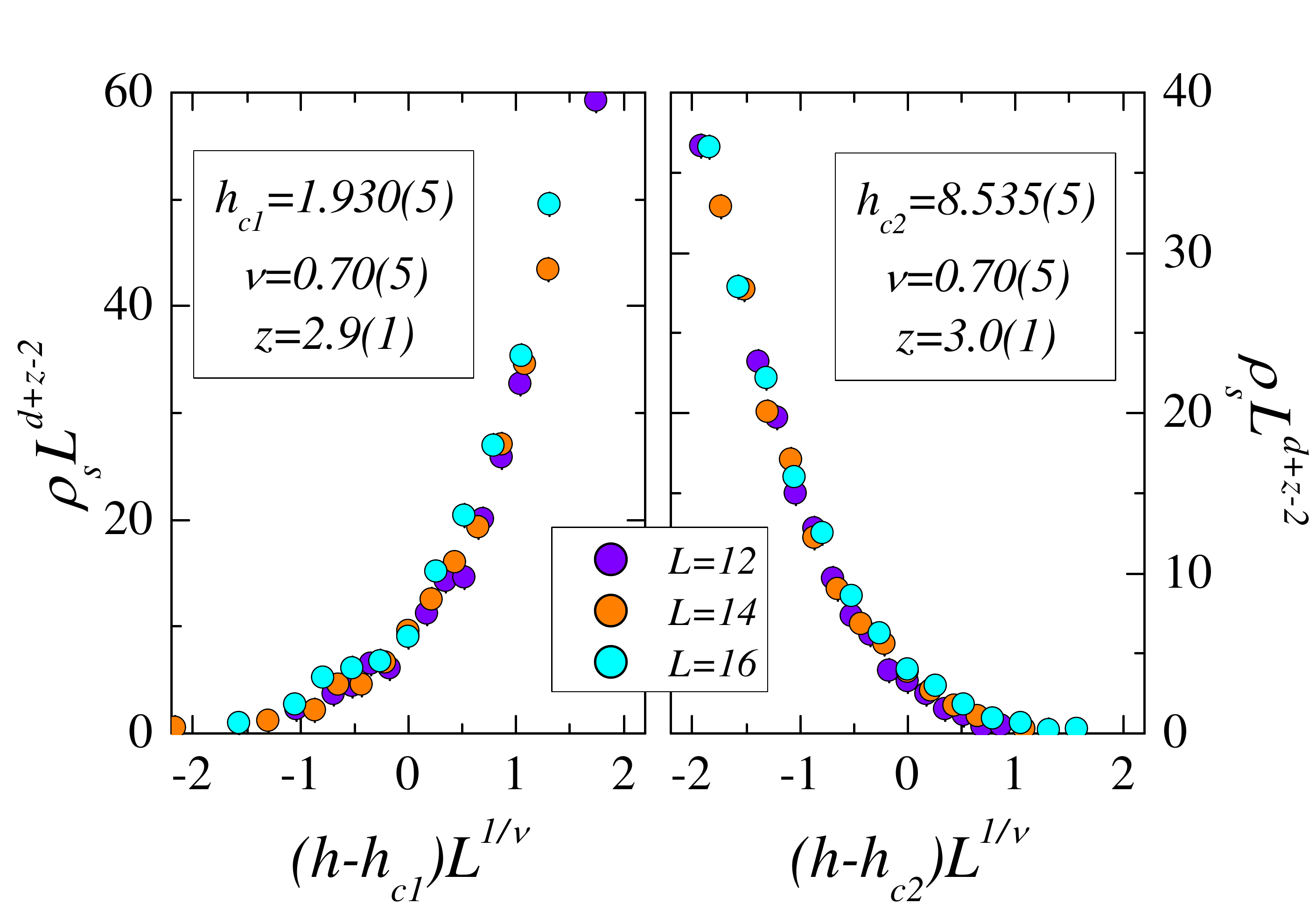}    
\caption{Scaling plots for the site-diluted DTN Hamiltonian with dilution $x=0.15$.}
\label{f.scalingsitedil}
\end{center}
\end{figure*}

\subsection{Site-diluted DTN}
 
We can repeat a similar analysis to that of Br-doped DTN for QMC results on site-diluted DTN with $x=0.15$ dilution. 
All results are summarized in the scaling plots of Fig.~\ref{f.scalingsitedil}, referring to the case of both $h_{c1}$ and $h_{c2}$. 
The central result is that all critical exponents appear to be consistent between the two transitions. In the case of site dilution, the asymmetry between the $ab$ plane and $c$ axis in terms of the disorder distribution is absent, and indeed we do not observe a significant discrepancy in the critical fields estimated via $\xi_{ab}$ and $\xi_c$; moreover we obtain a good collapse of the $\rho_s$ data, unlike what observed in Br-DTN. 

Even more importantly, the critical exponents estimated for site-diluted DTN are fully consistent with the estimates obtained above for Br-doped DTN, as well as with the estimates of Ref.~\onlinecite{HitchcockS06} for the Bose-Hubbard model with a random chemical potential. The latter reference gives $\nu = 0.70(12)$, $z=3$, and $\eta \approx - 1$, from which we
can extract $\beta = \nu(d+z-2+\eta)/2 \approx 1.05$. 
Hence we can conclude that none of the results obtained so far are specific of the kind of disorder we have considered, and that they rather reflect the universal behavior of interacting bosons in $d=3$ in the presence of short-range correlated disorder.

 \section{Thermodynamics along the quantum critical trajectory of Br-DTN: numerics and experiment}
 \label{s.QCtrajectory}
 
 A further test of the crossover scaling Ansatz comes from the direct determination of the
 exponents $x_C$ and $x_m$ from numerical simulations and experiments.  In the following
 we describe the extraction of these exponents from our quantum Monte Carlo simulations
 close to the $h_{c1}$ critical field for the Br-DTN Hamiltonian, as well as from specific heat experiments on Br-DTN. 
 
 \subsection{Quantum Monte Carlo}
   
 \begin{figure}[h]
\begin{center}
\includegraphics[
%bbllx=60pt,bblly=50pt,bburx=510pt,bbury=450pt,%
    width=90mm,angle=0]{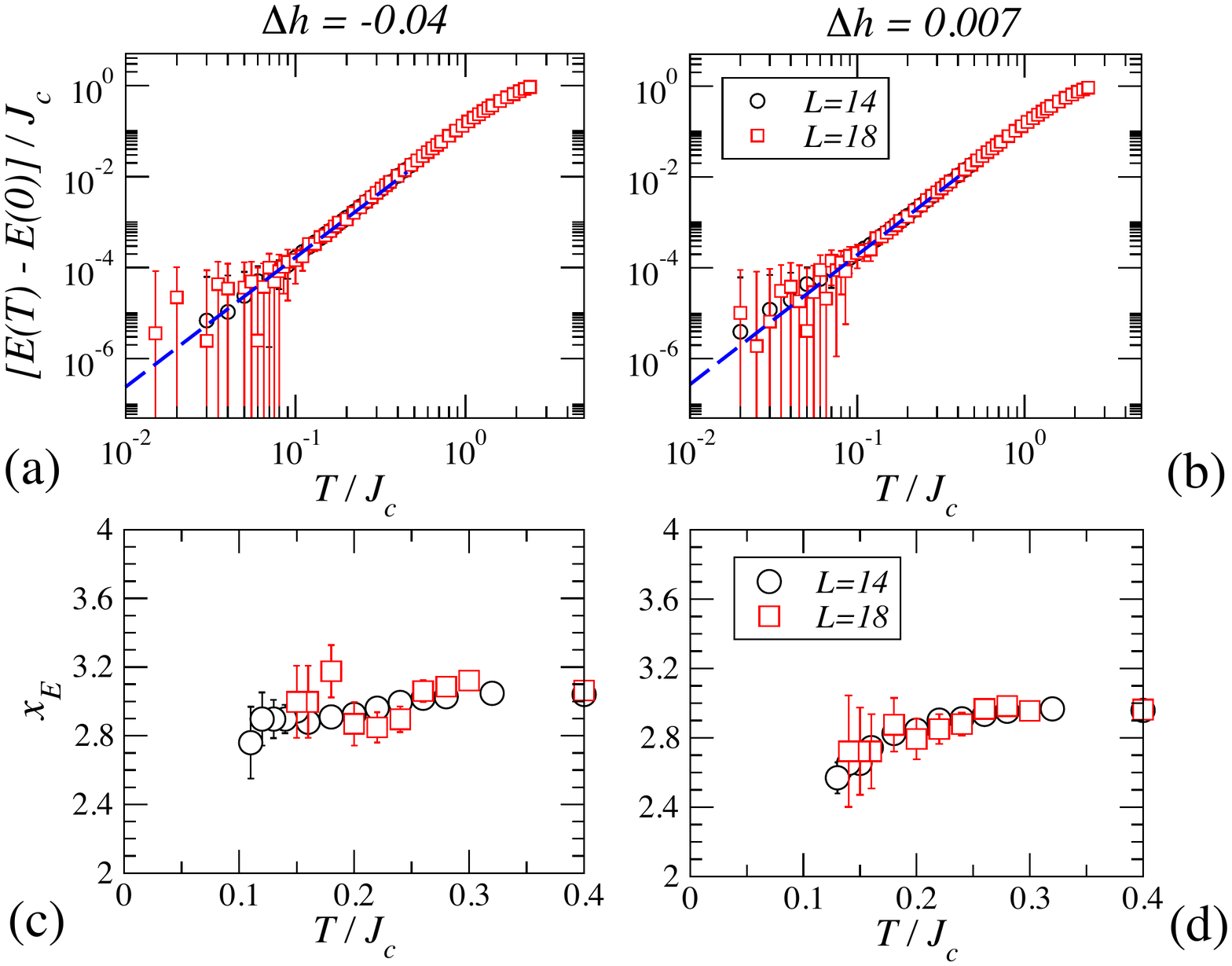}
\caption{(a)-(b) Thermal energy $E(T) - E(0)$  of the Br-DTN Hamiltonian close to the lower critical field $h_{c1}$.
The dashed lines correspond to a fit with $x_E = 2.85$;
(c)-(d) $x_E$ exponent obtained by fitting the data in (a)-(b) over a temperature window $[0,T]$.}
\label{f.dEwindow}
\end{center}
\end{figure}
   
   In Fig.~\ref{f.dEwindow} we analyze the scaling of the energy 
   per spin,  $E(T)$, as obtained from our QMC simulations. 
   The thermal energy is expected to scale as a power
   law close to the QCP, $E(T) = E(0) + A_E T^{x_E}$ where $x_E = x_C + 1$. 
  We fit our QMC data for two field values close to the estimated $h_{c1}=0.828(3)$ T
  value, $h_{c1} - 0.04$ T and $h_{c1} + 0.007$ T, and for two system sizes, $L=14$ and 
  $L=18$.
  We make use a windowing technique,   
  extracting the fitting parameters over a decreasing temperature window $[0,T]$
  in order to obtain the asymptotic quantum critical behavior for $T\to 0$.
  The results of this fitting procedure are shown in Fig.~\ref{f.dEwindow}. For both system sizes and 
  field values we obtain that the $x_E$ estimate converges to a value $x_E = 2.8(2)$
  as the temperature window for the fit is progressively reduced, whence we extract
  the estimate $x_C = 1.8(2)$. 
   
  \begin{figure}[h]
\begin{center}
\includegraphics[
%bbllx=60pt,bblly=50pt,bburx=510pt,bbury=450pt,%
    width=65mm,angle=0]{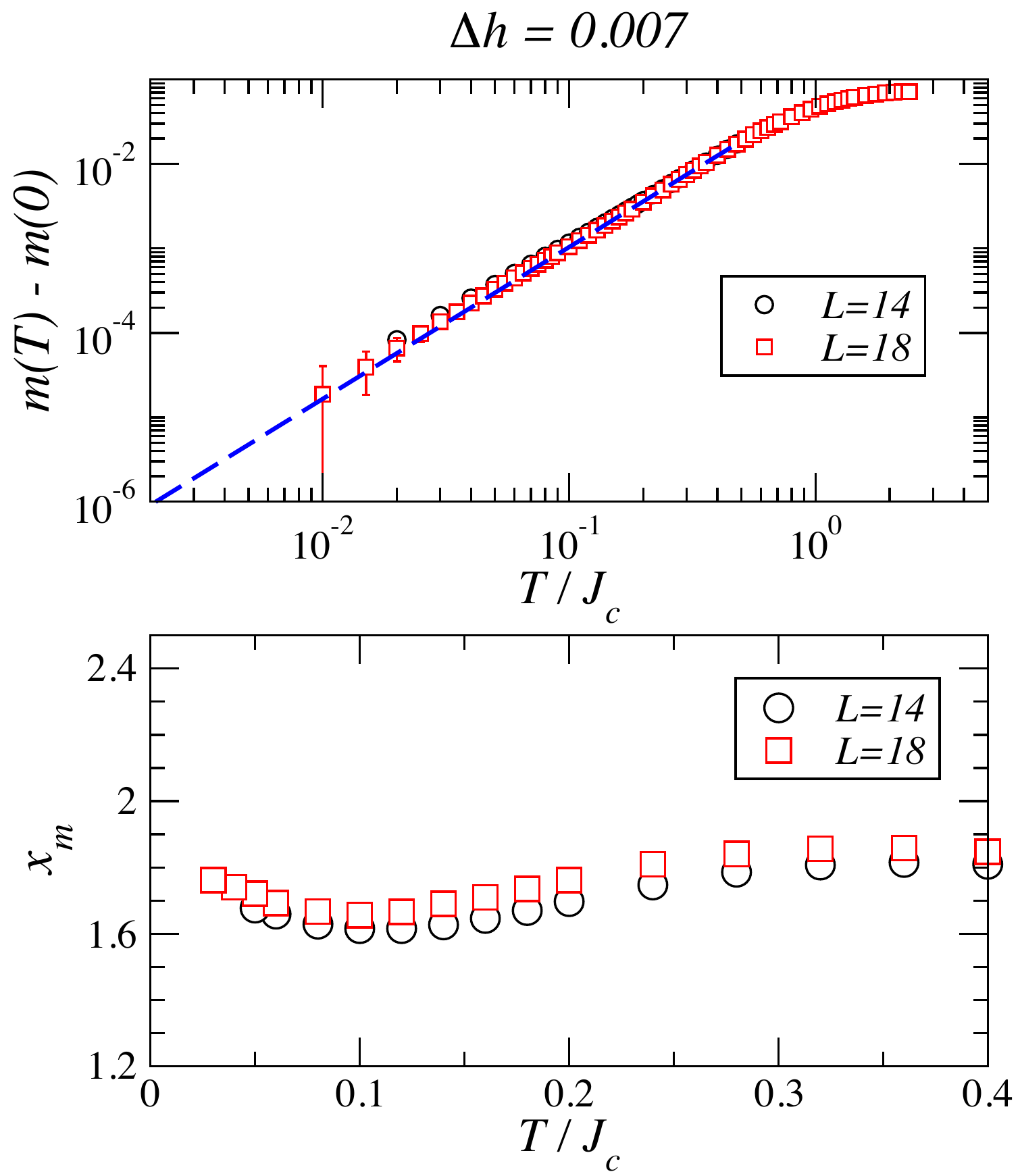}
\caption{Upper panel: thermal magnetization of the Br-DTN Hamiltonian close to the lower critical field $h_{c1}$; the
dashed line corresponds to a fit with $x_m = 1.8$. Lower panel: $x_m$ exponent obtained
by fitting the $m(T)$ data over a temperature window $[0,T]$.}
\label{f.mexp}
\end{center}
\end{figure} 
   
  \begin{figure}[h!]
\begin{center}
\includegraphics[
%bbllx=60pt,bblly=50pt,bburx=510pt,bbury=450pt,%
    width=90mm,angle=0]{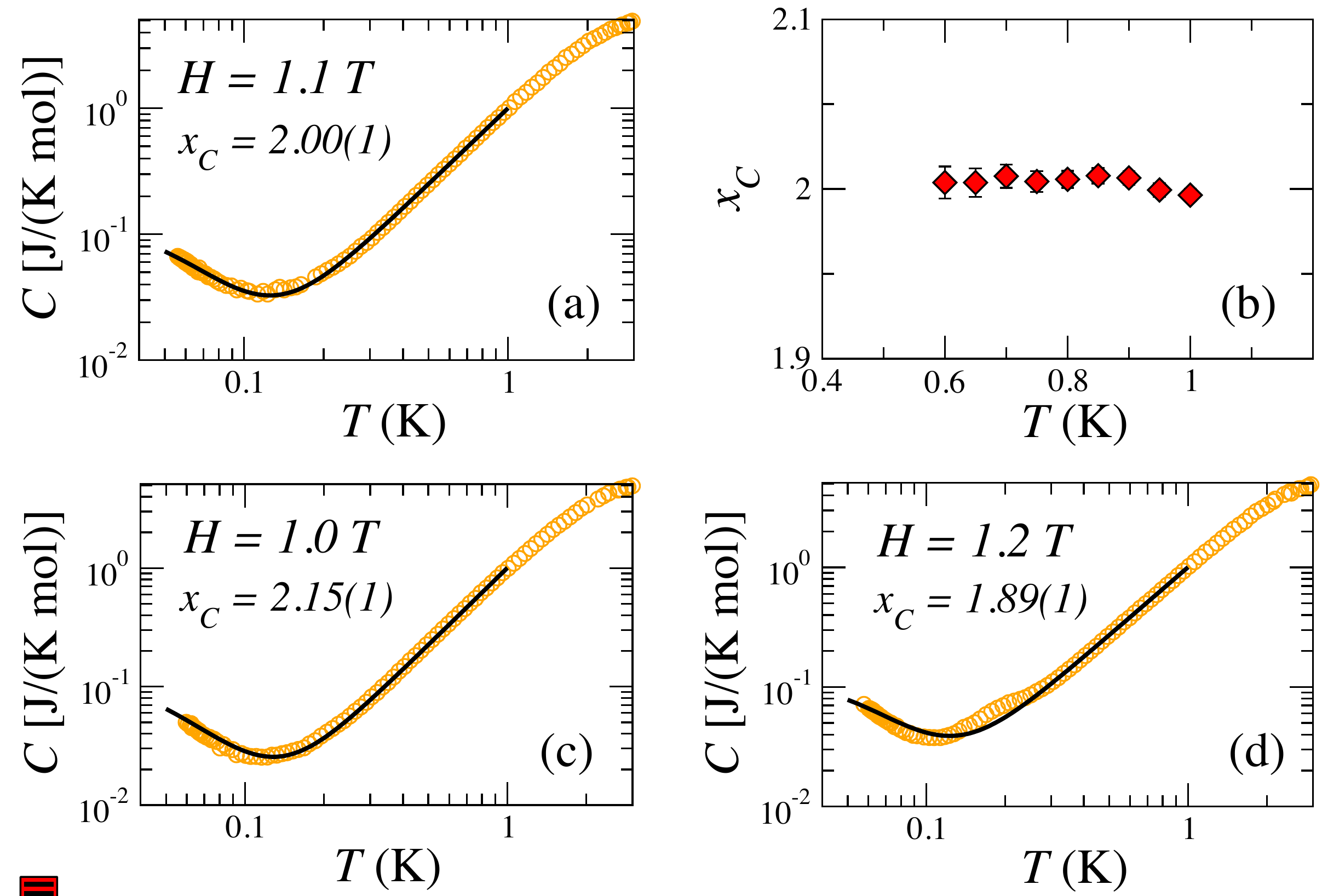}
\caption{Specific heat of Br-DTN close to the lower critical field $H_{c1}$.
The solid lines are fits as described in the text.}
\label{f.Cexp}
\end{center}
\end{figure} 

A similar procedure is used to analyze the magnetization
$m(T)$, which is fitted to the form $m(T) = m(0) + A_mT^{x_m}$
where $A_m$ is a constant. The result of the windowing analysis
is shown in Fig.~\ref{f.mexp}, yielding an exponent $x_m = 1.8(1)$. 
 
 \subsection{Specific heat measurements}
 
   We can perform a similar analysis to the experimental specific heat data of Br-doped DTN 
   close to the experimental $H_{c1}$ value, $H_{c1}= 1.07(1)$, namely for $H = 1, 1.1$
   and 1.2 T. The specific heat has been measured in a dilution refrigerator using the quasi-adiabatic heat pulse method.
   The analysis of the low-$T$ experimental data for the specific heat 
   is complicated by the appearance of a nuclear Schottky anomaly, which 
   introduces a term that scales roughly as $H^2/T^2$ in the specific heat - at least at sufficiently small fields - 
   as already observed in pure DTN.\cite{Kohamaetal11, Weickertetal12} The 
   Schottky anomaly contribution to the specific heat at $H_{c2}$ swamps the 
   low-$T$ electronic contribution, and thus we focus on data near $H_{c1}$. 
    
   We fit the specific heat data to the form 
   \begin{equation}
   C(T) = \left(\frac{a_n H} {T}\right)^2 \left[ 1 - \tanh^2 ( b_n H / T) \right] + a_e ~T^{x_C}
   \label{e.fit}
   \end{equation} 
   where the first term mimics the temperature behavior of the specific heat of paramagnetic nuclear spins, and the 
    second term is the electron-spin contribution. 
   In particular, we fit the specific heat data at $H = 1.1$ T over a variable temperature range [0,T] using the same windowing technique as for the theoretical data
   (see Fig.~\ref{f.Cexp}(b)).   
   This gives us stable fit parameters as $T$ is lowered, and for $T=0.6$ K we obtain a very good fit with $a_n =  0.0155(2)$ [JK/(T$^2$mol)]$^{1/2}$, $b_n = 0.034(2)$ K/T, $a_e =  1.004(7)$ J/(K$^{1+x_C}$ mol) and $x_C = 2.00(1)$ (see Fig.~\ref{f.Cexp}(a)).
   
   The other data sets are instead fit differently: we use as fitting parameters $a_e$ and $x_C$ only, while $a_n$ and $b_n$ are kept fixed - as the nuclear contribution depends explicitly on the field. For $H=1.0$ T we obtain $a_e = 1.03(1)$ J/(K$^{1+x_C}$ mol) and $x_C = 2.15(1)$, while for $H=1.2$ T we find $a_e = 1.004(1)$ J/(K$^{1+x_C}$ mol) and $x_C = 1.89(1)$ -- see Fig.~\ref{f.Cexp}(c-d). For both fields the nuclear part is quantitatively reproduced by using the explicit field dependence of Eq.~\eqref{e.fit}.   
   In particular for $H = 1.2$ T the experimental data deviate significantly from the fitting form for $T \approx 0.2$ K, due to the apparition of an anomaly associated with the onset of long-range order in the electronic spins (the value of $T_c \approx 0.2$ K at $H=1.2$ T is in good agreement with the $T_c(H)$ curve reported in Ref.~\onlinecite{Yuetal11}). Nonetheless above $T_c$ the specific heat appears to follow a power-law scaling characteristic of the quantum-critical behavior.

 \subsection{Discussion}
 
 The specific heat exhibits a power-law behavior along the quantum critical
 trajectory with an exponent $x_C = 1.8-2$ determined both by numerical simulations
 and by experiments on Br-DTN. Such an exponent is a clear signature of the 
 quantum critical behavior, as it clearly differs from the power-law behavior 
 with $x_C = 3$ appearing in the BEC phase \cite{Yuetal11}. 
 Moreover, this result is inconsistent with the prediction $x_C= 1$
 coming from the scaling Ansatz Eqs.~\eqref{e.onepscaling} and
 Eq.~\eqref{e.xc} with $z=d$.  
 The magnetization also exhibits a distinct quantum critical behavior with
 exponent $x_m \approx 1.8$. Such a behavior is markedly different 
 from that exhibited in the BEC regime, in which the thermal magnetization 
 initially decreases as $-T^{3/2}$ when $T$ grows above zero \cite{Nikunietal00}. 
  In turn the quantum critical exponent $x_m$ found with QMC is not consistent with the 
 prediction $x_m \approx 1.5$ from Eqs.~\eqref{e.onepscaling} 
 and Eq.~\eqref{e.xm} with $z=d$ and $\nu \approx 0.7$. 
 These results, together with the value $\phi = 1.1(1)$, call for a revision of the
 scaling behavior of the free energy governing the region around the QCP.
 
 \section{Generalized scaling Ansatzes}
 \label{s.ansatz}
 
  As already mentioned in Sec.~\ref{s.dbQCP}, the long-wavelength effective action describing the physics of dirty bosons
  around the QCP does not lend itself to a well controlled RG analysis, 
  due to the non-perturbative role of disorder, and to the fact that the 
  upper critical dimension appears to be $d_c = \infty$. \cite{Fisheretal89, Weichman08} 
  Therefore an educated guess of the scaling Ansatz for the free energy
  is not straightforward. In the following we discuss two attempts at a generalization
  of the scaling Ansatz, Eq.~\eqref{e.onepscaling}, based on the possible presence
  of a dangerously
  irrelevant term in the effective action, namely a direction in parameter space which cannot be
  simply eliminated in the RG flow. We cannot say a priori whether the term
  in question is represented by the interaction or by the disorder, or by a combination
  of the two. The generalization of the scaling Ansatz, although introduced in a heuristic way, can lead to 
  relationships between the critical exponents which are compatible with the numerical
  and experimental values cited above for Br-DTN. A fully consistent picture with our results can only be obtained 
  if a dangerously irrelevant operator enters in a \emph{weak} way in the scaling, only affecting the finite-temperature behavior.

 \subsection{Scaling Ansatz in presence of a dangerously irrelevant term}
 
 We consider the free energy density $f_s(g,T,u)$, depending on the distance $g$
 to the critical point, the temperature $T$ and the coefficient of a dangerously
irrelevant term $u$.
 Close to the QCP, $f_s$ is assumed to scale under $m$ cycles of an RG transformation -- scaling the 
spatial dimensions by a factor $b$ and the imaginary-time dimension by a factor
$b^{z}$ -- as 
\begin{equation}
f_s(g,T,u) \approx b^{-m(d+z)} f_s \left( g~ b^{m y_g}, T ~b^{m y_T}, u ~b^{-m y_u} \right)~.   
\end{equation} 
where $y_g$, $y_T$ and $y_u > 0$.
Stopping the RG flow when $g b^{m y_g} \approx 1$, namely when the correlation length is renormalized
to order 1, $\xi/b^m \sim 1$ with $y_g = 1/\nu$, and identifying $y_T = z$, one obtains the 
standard scaling Ansatz 
%\cite{...}
\begin{equation}
f_s(g,T,u) \approx g^{\nu(d+z)}~\Phi\left( \frac{T}{g^{\nu z}}, u ~g^{\nu y_u} \right)~.
\end{equation} 
Letting $u\to 0$ leads to singularities in $f_s$ or in its derivatives. 
Following Ref.~\onlinecite{Fisher83} one can assume that $\Phi(y_1,y_2)$ has a singularity
when $y_2 \to 0$, in the form
\begin{equation}
\Phi(y_1,y_2) \approx A(y_1 y_2^{\rho})~ y_2^{-\mu}
\end{equation}
where $\rho$ and $\mu$ are unknown exponents, except for the fact that $\mu > 0$. 
This assumption immediately leads to the scaling Ansatz
\begin{equation}
f_s(g,T,u) \approx g^{2-\alpha} A\left(\frac{T}{g^{\phi}} u^\rho \right)
\label{e.ansatz1}
\end{equation}
where $2-\alpha = \nu(d+z) - \nu y_u \mu \leq \nu(d+z)$ (in possible violation of hyperscaling)
and $\phi = \nu z - \rho y_u \nu \leq \nu z$ (if $\rho>0$). Analogously to what discussed in Sec.~\ref{s.crossover},
the thermal critical line must correspond to a singularity in $A(y)$ at $y_c$, so that 
$T_c \sim g^{\phi}$. The inequality $\phi \leq \nu z$ is clearly verified by 
our numerical and experimental values for the $\phi$, $\nu$ and $z$ exponents
of Br-DTN. 

 Rewriting Eq.~\eqref{e.ansatz1} as 
 \begin{equation}
 f_s(g,T,u) \approx T^{\frac{2-\alpha}{\phi}} u^{\frac{\rho(2-\alpha)}{\phi}} B\left( \frac{g}{T^{1/\phi} ~u^{\rho/\phi}} \right)
 \end{equation}
 and applying Eqs.~\eqref{e.Cscaling} and \eqref{e.mscaling}, one obtains the following exponents for the along the quantum critical trajectory
 \begin{equation}
 x_m = \frac{1-\alpha}{\phi} ~~~~~~~~~~ x_C = x_m + \frac{1}{\phi} - 1
 \label{e.xCxm}
 \end{equation}
 Due to the violation of hyperscaling, $\alpha$ is a priori unknown. One can determine
 $\alpha$ by imposing that $x_m = 1.8(1)$, obtaining $\alpha = -0.98(1)$, 
 and hence $x_C = 1.7(3)$, in good agreement with the value obtained numerically
 and experimentally for Br-DTN. 
 
 Yet the above result has the major drawback of requiring a strong violation of hyperscaling at the QPT, 
  $2-\alpha \leq \nu (z+d)$. Violation of hyperscaling is in open contradiction with the analysis of Sec.~\ref{s.zd}, which is successfully based on conventional scaling forms for a QPT below the upper critical dimension. This suggests that a possible dangerously irrelevant operator enters in a more subtle way, leaving the quantum critical behavior unaffected and only manifesting itself in the finite-temperature singularity of the free energy.

 \subsection{Continentino's scaling}
 
 The previous scaling Ansatz, albeit giving the prediction Eq.~\eqref{e.xCxm} which
 is fully compatible the exponents for Br-DTN, is not the most general Ansatz in presence
 of a dangerously irrelevant term (as also pointed out in Ref.~\onlinecite{Fisher83}).
 In the absence of disorder, the boson-boson interaction is known rigorously
 to play the role of a dangerously irrelevant term \cite{Fisheretal89}, and yet 
 Eq.~\eqref{e.xCxm} is not satisfied: indeed one has that $\phi = 2/3$ (see Ref.~\onlinecite{Nikunietal00}) and $\alpha=0$
 (Gaussian fixed point), leading to $x_C = 2$, while the exact result is $x_C = 3/2$, 
 as also measured experimentally in undoped DTN.\cite{Kohamaetal11} 
 
  Continentino \cite{Continentinobook} has proposed an alternative scaling Ansatz to describe
  the singular part of the free energy close to the classical critical line (namely in the 
  so-called Ginzburg region, see Fig.~\ref{f.QCcartoon}) in the presence of a dangerously
  irrelevant term:
  \begin{equation}
  f_s(g,T,u) \approx {\tilde g}(T)^{2-\alpha}~ \Psi \left( \frac{T}{{\tilde g}(T)^{\nu z}} \right)
  \label{e.Continentino}
  \end{equation} 
 where ${\tilde g}(T) = g - uT^{1/\phi}$ is the distance to the thermal critical value of
 the control parameter, shifted with respect to its $T=0$ value $g$.
 The function $\Psi(t)$ is supposed to have the following properties: 1) $\Psi(t\to 0) =$ const., so that the Ansatz reproduces the conventional 
 quantum critical behavior $f_s \sim g^{2-\alpha}$ at $T=0$; 2) $\Psi(t\to\infty) \sim t^x$, where
 $x = (\alpha_T - \alpha)/(\nu z)$, and $\alpha_T$ is the critical exponent of the thermal transition, so that
 $f_s \sim |\tilde g(T)|^{2-\alpha_T}$ when approaching the thermal critical line. 
 In particular in the above Ansatz the operator $u$ \emph{disappears} from the scaling when $T\to 0$, leaving the QPT unaffected. 
 Hence hyperscaling can be safely assumed at $T=0$. 
  
  In general Eq.~\eqref{e.Continentino} is only guaranteed to work in the Ginzburg region
  \cite{Zhuetal04}, but one can test the consequences of its extension 
  to a broader range of validity, including the quantum critical trajectory, as done in 
Ref.~\onlinecite{Continentinobook}. Using hyperscaling at both the classical and quantum phase transition, 
this leads to the predictions
\begin{equation}
x_C = \nu_T d \left( \frac{1}{\phi} - \frac{1}{\nu z} \right) + \frac{d}{z}
\label{e.xcCont}
\end{equation}
and 
\begin{equation}
x_m = x_C + 1 - \frac{1}{\phi}
\end{equation}
where $\nu_T$ is the critical exponent at the thermal transition. 

Using $\phi=1.1(1)$, $\nu=0.70(5)$, $z=d$ and $\nu_T = 0.669(1)$ for the 3DXY universality class \cite{PelissettoV02}, 
we obtain $x_C = 1.9(2)$ and $x_m = 2.0(2)$, fully compatible with our numerically and experimentally determined values. 

A possible argument can be formulated to justify the apparent success of the above Ansatz \emph{a posteriori}.
This Ansatz implies the rather unconventional situation, Eq.~\eqref{e.xcCont},  in which both the
quantum and the classical critical exponents enter in the scaling along the quantum critical trajectory \cite{Zhuetal04}.
This is not what one would typically expect, given that the classical critical exponents pertain to the Ginzburg region, while the quantum critical 
exponent pertain to the quantum critical fan above the QCP. The Ginzburg region surrounds the classical critical line $T_c \sim |g|^{\phi}$, while the quantum critical fan is 
(loosely) lower bounded by a crossover line $T_{\rm co} \sim g^{\nu z}$ reflecting the typical energy scale of the spectral features which vanish at the QCP.   
If $\phi = \nu z$, as predicted by conventional scaling, Eq.~\eqref{e.onepscaling}, the two above regions can be well separated (Fig.~\ref{f.QCcartoon}(a)). On the other hand, if 
$\phi < \nu z$ (as in the dirty-boson case, in which $\nu z \approx 2 \phi$) the quantum critical regime is not lower-bounded by $T_{\rm co}$ but rather by $T_c$, and hence the quantum-critical region and the Ginzburg region are completely contiguous, so that one could imagine that the quantum critical trajectory could be influenced by the classical critical behavior (Fig.~\ref{f.QCcartoon}(b)). This very unconventional scenario is purely speculative at this stage, and it will require more investigations \cite{remark}.   
     
 \subsection{Alternative scaling}
     
 To avoid the shortcomings of Continentino's scaling - namely the extension to the quantum-critical trajectory of a scaling form expected to be valid close to the classical critical line only - we can propose a further scaling Ansatz which generalizes Continentino's one while preserving its virtues. Indeed Eq.~\ref{e.Continentino} is consistent with the more general, \emph{two-argument} scaling Ansatz
 \begin{equation}
  f_s(T,g,u) \approx  g^{\nu(d+z)} ~{\cal G} \left( \frac{g}{T^{1/(\nu z)}}, u ~T^{\frac{1}{\phi}- \frac{1}{\nu z}}\right)~.
  \label{e.generalAnsatz}
  \end{equation} 
 To obtain hyperscaling at $T=0$, we simply need to assume that ${\cal G} (y_1, y_2) \to$ const. when 
 $y_1 \to \infty$ and $y_2 \to 0$.  

We rewrite the above scaling form as  
 \begin{equation}
  f_s(T,g,u) \approx  T^{\frac{d+z}{z}} ~{\cal F} \left( \frac{g}{T^{1/(\nu z)}}, u ~T^{\frac{1}{\phi}- \frac{1}{\nu z}}\right)
  \label{e.generalAnsatz2}
  \end{equation} 
  where ${\cal F}  \to (g/T^{1/(\nu z)})^{\nu(d+z)} {\cal G}$ when approaching the QCP at very low temperatures, namely $T\to 0$ and $g\to 0$ but $g/T^{1/(\nu z)} \to \infty$ (implying $T \ll T_{\rm co} \sim g^{\nu z}$).
 
  On the other hand, when approaching the quantum-critical trajectory, $T\to 0$, $g\to 0$, but $g/T^{1/(\nu z)} \to 0$
  (namely $T \gg T_{\rm co} \sim g^{\nu z}$), we assume a rather different form for the ${\cal F}$ function, namely 
  \begin{equation}
  {\cal F}(y_1, y_2) \approx y_2^{\theta} (f_1 y_1 + f_2 y_2 + ....)
  \label{e.Ffunction} 
  \end{equation}
  when $y_1, y_2 \to 0$ ($f_1$ and $f_2$ are constants) \cite{remark2}. Upon this assumption we readily obtain that, along the quantum-critical trajectory $y_1 = 0$, $y_2 \to 0$
  \begin{equation}
  x_C = \frac{d+z}{z} + (\theta+1) \left( \frac{1}{\phi} - \frac{1}{\nu z} \right) - 1
  \label{e.xCg}
  \end{equation}
 and 
 \begin{equation}
 x_m = x_C - \frac{1}{\phi} + 1~.
 \end{equation} 
 The $\theta$ exponent can be determined by Eq.~\eqref{e.xCg}, $\theta \approx 0.9$, using the numerical values of all the other exponents. Hence we can obtain a picture which is globally consistent with all of our results.

 \section{Conclusions}  
 \label{s.conclusions}
 
 We have investigated, both theoretically and experimentally, the field-induced quantum phase transition in the magnetic Hamiltonian of DTN in the presence of 
 bond/anisotropy disorder or of site dilution. This quantum phase transition represents a well controlled realization of the 
 dirty-boson quantum critical point (QCP) in three dimensions. 
 We have provided compelling evidence that conventional scaling is obeyed at the dirty-boson QCP, but that unconventional scaling (including the possible presence of a dangerously irrelevant operator) is necessary to account for the thermodynamics along the quantum critical trajectory, as well as for the power-law dependence of the line of critical temperatures on the applied field. Our results are compatible with the prediction $z=d$, but incompatible with the prediction $\phi = \nu z$ for the exponent of the power-law scaling of the critical temperature, as also found in recent experiments \cite{Yuetal11, Yamadaetal11, ZheludevH11, Yamadaetal11-2, Huvonenetal12}. Within a generalized scaling Ansatz $\phi$ plays the role of an independent exponent, whose value, combined with that of quantum critical exponents as well as classical critical ones, can be quantitatively related to the exponents governing the power-law scaling of the specific heat and of the magnetization along the quantum critical trajectory. A recent experiment on piperazinium-Cu$_2$(Cl$_{1-x}$Br$_x$)$_6$ has measured the order-parameter exponent via neutron scattering at low temperature,\cite{Huvonenetal12} giving $\beta \approx 0.5$. Our evidence is that this exponent is not compatible with the dirty-boson universality class in $d=3$ (see also Ref.~\onlinecite{HitchcockS06}); more investigations will be necessary in this direction.   
 
 While a conclusive picture on the dirty-boson quantum critical point is still far from obvious, our results provide a quantitative scenario which a well-controlled theory of the dirty-boson effective action \cite{Fisheretal89} should be able to reproduce. More generally, our results further confirm that disordered quantum magnets represent an invaluable tool for the investigation of complex phenomena of interacting bosons, and that they stand among the best candidate systems to unveil the universal features of dirty bosons.

\section{Acknowledgements}
We thank Qimiao Si for bringing Ref.~\onlinecite{Zhuetal04} to our
attention. R. Y and T. R. acknowledge support of the DOE (INCITE award).


\begin{thebibliography}{99}
\bibitem{Crowelletal97} P. A. Crowell, F. W. Van Keuls, and J. D. Reppy, Phys. Rev. B {\bf 55}, 12620 (1997).
\bibitem{SanchezPalenciaL10} L. Sanchez-Palencia and M. Lewenstein, Nature Phys. {\bf 6}, 87 (2010).
\bibitem{Hongetal10}
T. Hong, A. Zheludev, H. Manaka, and L.-P. Regnault,
Phys. Rev. B {\bf 81}, 060410(R) (2010).
\bibitem{Yamadaetal11} F. Yamada, H. Tanaka, T. Ono, and H. Nojiri,
Phys. Rev. B {\bf 83}, 020409 (2011). 

\bibitem{Yuetal11} R. Yu, L. Yin, N. S. Sullivan, J. S. Xia, C. Huan, A. Paduan-Filho, N. F. Oliveira Jr., S. Haas, A. Steppke, C. F. Miclea, 
F. Weickert, R. Movshovich, E.-D. Mun, B. S. Scott, V. S. Zapf, T. Roscilde, arXiv:1109.4403 (2011).
\bibitem{Huvonenetal12} D. H\"uvonen, S. Zhao, M. M\aa nsson, T. Yankova, E. Ressouche, C. Niedermayer, M. Laver, S. N. Gvasaliya, and A. Zheludev, 
arXiv:1201.6143 (2012).
\bibitem{Weichman08} P. B. Weichman, Mod. Phys. Lett. B  {\bf 22}, 2623 (2008).
 \bibitem{Fisheretal89} M. P. A. Fisher, P. B. Weichman, G. Grinstein, and D. S. Fisher, Phys. Rev. B {\bf 40}, 
546 (1989). 
\bibitem{WeichmanM07} P. B. Weichman and R. Mukhopadhyay, Phys. Rev. Lett. {\bf 98}, 245701 (2007).
\bibitem{Griffin09} A. Griffin, T. Nikuni, and E. Zaremba, \emph{Bose-condensed gases at finite temperatures}, Cambridge, 2009.
\bibitem{Sachdev99} S. Sachdev, \emph{Quantum phase transitions}, Cambridge, 1999.
\bibitem{Nikunietal00} T. Nikuni, M. Oshikawa, A. Oosawa, and H. Tanaka,
Phys. Rev. Lett. {\bf 84}, 5868 (2000).
\bibitem{GiamarchiS88} T. Giamarchi and H. J. Schulz, Phys. Rev. B {\bf 37}, 325 (1988).
\bibitem{Parisi12} G. Parisi, arXiv:1201.5813 (2012).
\bibitem{Priyadarsheeetal06} A. Priyadarshee, S. Chandrasekharan, J.-W. Lee, and H. U. Baranger, Phys. Rev. Lett. {\bf 97}, 115703 (2006).
\bibitem{MeierW12} H. Meier and M. Wallin, Phys. Rev. Lett. 108, 055701 (2012).
\bibitem{ProkofevS04} N. Prokof'ev and B. Svistunov, Phys. Rev. Lett. {\bf 92}, 015703 (2004).
\bibitem{Soyleretal11} S. G. S\"oyler, M. Kiselev, N. V. Prokof'ev, and B. V. Svistunov, Phys. Rev. Lett. {\bf 107}, 185301 (2011).
\bibitem{Linetal11} F. Lin, E. S\o rensen, and D. M. Ceperley, Phys. Rev. B {\bf 84}, 094507 (2011)
\bibitem{Roscilde06} T. Roscilde, Phys. Rev. B {\bf 74}, 144418 (2006).
\bibitem{Yuetal08} R. Yu, T. Roscilde and S. Haas, New J. Phys.
{\bf 10} 013034 (2008). 
\bibitem{Yuetal10} R. Yu, O. Nohadani, S. Haas, and T. Roscilde,  Phys. Rev. B {\bf 82}, 134437 (2010).
\bibitem{HitchcockS06} P. Hitchcock and E. S. S\o rensen, Phys. Rev. B {\bf 73}, 174523 (2006).
\bibitem{Giamarchi04} T. Giamarchi, \emph{Quantum physics in one dimension}, Clarendon, Oxford (2004).  
\bibitem{Wangetal06} L. Wang, K. S. D. Beach, and A. W. Sandvik, Phys. Rev. B {\bf 73}, 014431 (2006).
\bibitem{Nohadanietal05} O. Nohadani, S. Wessel, and S. Haas, Phys. Rev. B {\bf 72}, 024440 (2005).

\bibitem{AletS04} F. Alet and E. S. S\o rensen, Phys. Rev. B {\bf 70}, 024513 (2004). 
\bibitem{Giamarchietal08} T. Giamarchi,  C. R\"uegg, and O. Tchernyshyov,
Nature Phys. {\bf 4}, 198  (2008).
\bibitem{RoscildeH05} T. Roscilde and S. Haas,
Phys. Rev. Lett. {\bf 95}, 207206 (2005).
\bibitem{Nohadanietal05-2} O. Nohadani, S. Wessel, and S. Haas,
Phys. Rev. Lett. {\bf 95}, 227201 (2005).
\bibitem{ZheludevH11} A. Zheludev and D. H\"uvonen, Phys. Rev. B {\bf 83}, 216401 (2011).
\bibitem{Yamadaetal11-2} F. Yamada, H. Tanaka, T. Ono, and H. Nojiri,
Phys. Rev. B {\bf 83}, 216402 (2011). 


\bibitem{Zvyaginetal07}
S. A. Zvyagin, J. Wosnitza, C. D. Batista, M. Tsukamoto, N. Kawashima, J. Krzystek, V. S. Zapf, M. Jaime, N. F. Oliveira, Jr., and A. Paduan-Filho, 
Phys. Rev. Lett. {\bf 98}, 047205 (2007).

\bibitem{Yinetal08} L. Yin, J. S. Xia, V. S. Zapf, N. S. Sullivan, and A. Paduan-Filho
Phys. Rev. Lett. {\bf 101}, 187205 (2008).
 
\bibitem{Yuetal10-2} R. Yu, S. Haas, and T. Roscilde, 
Europhys. Lett. {\bf 89}, 10009 (2010).
\bibitem{Yinetal11} L. Yin, J.S. Xia, N.S. Sullivan, V. S. Zapf, A. Paduan-Filho,  R. Yu, and T. Roscilde, Proceedings of the LT26 Conference, to appear on 
J. Phys: Conf. Ser.
 \bibitem{Cardybook} J. Cardy, 
 \emph{Scaling and Renormalization in Statistical Physics}, Cambridge, 1996, Chap. 4.
\bibitem{Chayesetal86} J. T. Chayes, L. Chayes, D. S. Fisher, and T. Spencer, Phys. Rev. Lett. {\bf 57}, 2999 (1986).
\bibitem{SSE} O. F. Sylju\aa sen and A. W.  Sandvik, Phys. Rev. E {\bf 66}, 046701 (2002).
\bibitem{PollockC87} E. L. Pollock and D. C. Ceperley,
Phys. Rev. B {\bf 36}, 8343 (1987).
\bibitem{Sandvik02} A. W. Sandvik, Phys. Rev. B {\bf 66}, 024418 (2002).
\bibitem{Kohamaetal11} Y. Kohama, A. V. Sologubenko, N. R. Dilley, V. S. Zapf, M. Jaime, J. A. Mydosh, A. Paduan-Filho, 
K. A. Al-Hassanieh, P. Sengupta, S. Gangadharaiah, A. L. Chernyshev, and C. D. Batista,  
Phys. Rev. Lett. {\bf 106}, 037203 (2011).
\bibitem{Weickertetal12} F. Weickert, R. Kuechler, A. Steppke, L. Pedrero, M. Nicklas, M. Brando, F. Steglich, M. Jaime, V. S. Zapf, A. Paduan-Filho, K. A. Al-Hassanieh, C. D. Batista, and P. Sengupta, arXiv:1204.3064 (2012).
\bibitem{Fisher83} M. E. Fisher in \emph{Critical Phenomena}, Lecture Notes in Physics 186, Springer, Berlin (1983).
 \bibitem{Continentinobook} M. A. Continentino, \emph{Quantum Scaling in Many-Body
 Systems}, World Scientific, Singapore, 2001.
\bibitem{Zhuetal04} L. Zhu, M. Garst, A. Rosch, and Q. Si, arXiv:cond-mat/0408230.
\bibitem{PelissettoV02} A. Pelissetto and E. Vicari, Phys. Rep. {\bf 368}, 549 (2002).
 \bibitem{remark} Indeed a similar picture is \emph{a priori} obtained in the pure system, in which $\nu z = 1 > \phi = 2/3$. Nonetheless Continentino's scaling does not apply to the Gaussian quantum critical point of the pure system, as it predicts \emph{e.g.} the wrong $x_C$ exponent. 
 \bibitem{remark2} One can reconcile the two limiting cases of Eq.~\eqref{e.generalAnsatz} and Eq.~\eqref{e.Ffunction} by assuming that $ {\cal F}$ is a linear combination of the two expressions, given that the first will dominate over the second when approaching the QCP from low temperatures, while the second one will dominate close to the quantum-critical trajectory. On the other hand, the behavior in the Ginzburg region \emph{\`a la} Continentino can be reproduced by the ${\cal G}$ function term, while the contribution of Eq.~\eqref{e.Ffunction} in that region would be marginal.  

\end{thebibliography}
\end{document}